\documentclass[a4paper,fleqn,usenatbib]{mnras}
\pdfoutput=1
\usepackage{newtxtext,newtxmath}
\usepackage[T1]{fontenc}
\usepackage{ae,aecompl}

\usepackage{graphicx}
\usepackage{amsmath,amssymb}
\usepackage{gensymb}

\newcommand{\civ}{\hbox{C\,{\sc iv}}}
\newcommand{\oiii}{\hbox{O\,{\sc iii}}}
\newcommand{\oii}{\hbox{O\,{\sc ii}}}

\newcommand{\nv}{\hbox{N\,{\sc v}}}

\newcommand{\hii}{\hbox{H\,{\sc ii}}}
\newcommand{\hi}{\hbox{H\,{\sc i}}}
\newcommand{\heii}{\hbox{He\,{\sc ii}}}
\newcommand{\hei}{\hbox{He\,{\sc i}}}

\newcommand{\feiii}{\hbox{Fe\,{\sc iii}}}
\newcommand{\ariv}{\hbox{Ar\,{\sc iv}}}

\newcommand{\uband}{\ensuremath{u'}}
\newcommand{\gband}{\ensuremath{g'}}
\newcommand{\rband}{\ensuremath{r'}}
\newcommand{\iband}{\ensuremath{i'}}
\newcommand{\zband}{\ensuremath{z'}}

\newcommand{\unit}[1]{\ensuremath{\, \mathrm{#1}}}

\newcommand{\ott}{\ensuremath{\mathrm{O}_{32}}}
\newcommand{\rtt}{\ensuremath{\mathrm{R}_{23}}}

\title[New XMPs hosting young stellar populations]{Photometric identification and MMT spectroscopy of new extremely metal-poor galaxies: towards a better understanding of young stellar populations at low metallicity}
\author[P. Senchyna et al.]{
    Peter Senchyna \&
    Daniel P. Stark
    \vspace{0.1in}\\
    Steward Observatory, University of Arizona, 933 N Cherry Ave, Tucson, AZ 85721 USA \\
}

\date{Accepted XXX. Received YYY; in original form ZZZ}

\pubyear{2019}

\begin{document}
\label{firstpage}
\pagerange{\pageref{firstpage}--\pageref{lastpage}}
\maketitle

\begin{abstract}
    Extremely metal-poor star-forming galaxies (XMPs) represent one of our only laboratories for study of the low-metallicity stars we expect to encounter at early epochs.
    But as our understanding of the $z>6$ universe has improved, it has become clear that the majority of known XMPs within 100 Mpc host significantly less prominent massive star populations than their reionization-era counterparts, severely limiting their utility as testbeds for interpreting spectral features found at the highest redshifts.
    Here we present a new photometric selection technique designed to identify nearby XMPs dominated by young stellar populations comparable to those expected in the reionization era.
    We apply our technique to uncover candidate XMPs in SDSS imaging at magnitudes $16<\iband{}<23$, extending significantly below the completeness limits of the SDSS spectroscopic survey.
    Spectroscopic observations with the MMT confirm that 32 of the 53 uniformly metal-poor and high specific star formation rate targets we observed have gas-phase oxygen abundances $12+\log\mathrm{O/H}<7.7$ ($Z/Z_\odot<0.1$), including two in the range of the lowest-metallicity galaxies known, $Z/Z_\odot<0.05$.
    Our observations shed new light onto the longstanding mystery of \heii{} emission in star-forming galaxies: we find that the equivalent width of the \heii{} $\lambda 4686$ high-ionization emission line does not scale with that of H$\beta$ in our sample, suggesting that binary evolution or other processes on $>10$ Myr timescales contribute substantially to the $\mathrm{He^+}$-ionizing photon budget in this metallicity regime.
    Applying such selection techniques coupled with deep spectroscopy to next-generation photometric surveys like LSST may eventually provide a basis for an empirical understanding of metal-poor massive stars.
\end{abstract}


\footnotetext[1]{E-mail: senchp@email.arizona.edu}

\section{Introduction}

Blue compact dwarf galaxies host the nearest populations of very young stellar populations at low metallicities ($Z/Z_\odot<0.1$).
The spectra of these objects were found early-on to be strikingly similar to those of \hii{} regions \citep{Zwicky1966}, indicating a preponderance of hot stars ionizing their surrounding gas.
Since their discovery and the recognition of their significantly low metal content \citep[][]{Searle1972}, these systems have become fruitful laboratories for a wide range of astrophysics.
They have found a role in topics ranging from galactic evolution and chemical enrichment \citep[e.g.][]{Ekta2010,Izotov2012,Guseva2017} to measurement of the primordial helium abundance \citep[e.g.][]{Izotov1994,Stasinska2001,Cooke2015}.

Local star-forming dwarf galaxies have also proved crucial to understanding the earliest galaxies.
In recent years, detections of surprisingly strong emission in rest-ultraviolet (UV) nebular lines (in particular \civ{} and \nv{}) at reionization era redshifts have highlighted uncertainties in our models of stellar populations and their ionizing spectra below SMC metallicity \citep[e.g.][]{Stark2015a,Tilvi2016,Mainali2017,Laporte2017,Mainali2018}.
Deep ultraviolet and optical spectra of nearby star-forming galaxies suggest that the intense nebular \civ{} emission may be powered by very metal-poor stellar populations ($Z/Z_\odot<0.1$), and demonstrate that the extreme-UV (EUV) ionizing spectra predictions made by current models require substantial revision at low metallicities \citep{Senchyna2017}.
The accuracy of these stellar population synthesis models is also critical for understanding the early heating of the IGM prior to reionization, which is likely dominated by the first X-ray binaries in the universe \citep[e.g.][]{Fragos2013,Mirocha2015}.
X-ray observations of very metal-poor nearby galaxies suggest that stellar populations below $Z/Z_\odot < 0.1$ are far more efficient producers of hard X-ray radiation than those near solar \citep{Brorby2014,Douna2015}, with significant consequences for the early IGM and the global 21-cm signal produced in the early universe \citep[e.g.][and references therein]{Furlanetto2006}.

However, successfully employing local star-forming galaxies as reference objects in this context hinges upon identifying systems with properties comparable to reionization-era galaxies.
In the last decade, a series of deep imaging campaigns have revolutionized our view of early star-forming galaxies (see \citealt{Stark2016} for a review).
These surveys have revealed a population of compact and low-mass systems which are experiencing a substantial recent upturn in star formation rate (SFR).
This is reflected in very high specific star formation rates ($\mathrm{sSFR} = \mathrm{SFR/M_\star}\gtrsim $ 5--10 $\mathrm{Gyr^{-1}}$) and correspondingly high equivalent width line emission likely powered by dense clusters of hot metal-poor stars \citep[e.g.][]{Smit2015,Roberts-Borsani2016,Stark2015a,Mainali2017}.

Unfortunately, most extremely metal-poor galaxies (XMPs, with $Z/Z_\odot<0.1$ i.e.\ $12+\log\mathrm{O/H}<7.7$; \citealt{Asplund2009}) known in the local universe are characterized by significantly lower sSFRs than these reionization-era systems.
The set of relatively bright XMPs uncovered by the SDSS spectroscopic survey \citep[e.g.\ those assembled by][]{SanchezAlmeida2016} present a median H$\beta$ equivalent width of 30 \AA{} corresponding to a typical sSFR $\lesssim 1\; \mathrm{Gyr^{-1}}$, 5--10 times lower than systems at $z>6$ \citep[as inferred from SED fitting incorporating rest-optical nebular line contamination, e.g.][]{Salmon2015,Smit2015}.
\footnote{The equivalent width of H$\beta$ is a commonly-used empirical proxy for sSFR, or equivalently the light-weighted age of a stellar population, as it probes the ratio of H-ionizing flux dominated by very young ($<10$ Myr) massive stars to optical continuum emission typically dominated by longer-lived A stars. For simplicity and consistency with studies at $z>6$ \citep[for example, see discussion in][]{Salmon2015}, we assume a constant star formation history in converting between H$\beta$ equivalent widths and inferred sSFR.}
As a result, the X-ray and nebular line spectra of typical known XMPs likely have significant contributions from somewhat older populations, including the energetic products of binary evolution involving stars with main sequence lifetimes of 10--100 Myr \citep[e.g.][]{Fabbiano2006,Eldridge2017,Gotberg2017}.
Using these systems as laboratories in which to study the properties of very young stellar populations at reionization-era metallicities may then lead to significant inaccuracies in our understanding of early galaxies and IGM heating.

Motivated by the paucity of extremely metal-poor systems at sSFRs comparable to those found in the reionization era, we have begun a campaign to identify new XMPs dominated by very young stellar populations within 100 Mpc ($z<0.03$).
We developed a set of photometric color cuts designed to select these objects from $ugriz$ photometry based upon the colors of systems like I Zw 18 with significant contamination of both the $g$ and $r$ bands by nebular emission from [\oiii{}]+H$\beta$ and H$\alpha$.
In contrast to previous photometric excess selections \citep[e.g. the Green Peas and Blueberries][]{Cardamone2009,Amorin2010,Yang2017}, our two-band excess method targets relatively low [\oiii{}]/H$\alpha$ ratios as observed in very low-metallicity systems.
By extending this search to objects several magnitudes fainter than SDSS spectra access and focusing on very low redshifts, we aim to identify lower-mass and lower-metallicity systems missed by large spectroscopic surveys, a number of which have been uncovered previously \citep[e.g.][]{James2015,Hirschauer2016,SanchezAlmeida2017,Hsyu2017}.
Developing and testing such photometric selection techniques is particularly pressing in light of ongoing and imminent deep photometric surveys such as that to be conducted by the Large Synoptic Survey Telescope \citep[LSST;][]{Ivezic2008}.

Here, we present the results of MMT spectroscopic follow-up of objects selected using our technique from SDSS photometry at magnitudes $18<i<21$.
We present the photometric selection and sample in Section~\ref{sec:photsamp}.
In Section~\ref{sec:obsanalysis} we describe the photometric data, spectroscopic follow-up, and analysis.
We present the results in Section~\ref{sec:results}, and conclude in Section~\ref{sec:conclude}.
We assume a solar oxygen abundance of $12+\log_{10}\left([\text{O/H}]_\odot\right)  = 8.69$ \citep{Asplund2009}.
For distance calculations and related quantities, we adopt a flat cosmology with $H_0 = 70$ \unit{km \, s^{-1} \, Mpc^{-1}}.
All magnitudes are in the AB system unless otherwise stated.

\section{Photometric Sample}
\label{sec:photsamp}

We aim to identify XMPs from ground-based broadband photometric data.
In particular, we focus on the commonly-used Sloan photometric system, which consists of five wide bands denoted \uband{}\gband{}\rband{}\iband{}\zband{} spanning the wavelength range 3000-11000 \AA{} \citep{Fukugita1996}.
The SDSS employed this system in imaging spanning over 11,663 \unit{deg^2} of the Northern sky to a depth of $r= 22.2$ ($z=20.5$), with accompanying fiber spectroscopy complete for only the brightest galaxies $r<17.77$ \citep{York2000,Strauss2002,Abazajian2009}.
Ongoing large surveys such as the Hyper Suprime-Cam Subaru Strategic Program \citep[HSC-SSP;][]{Aihara2018} and Dark Energy Survey \citep[DES;][]{Abbott2018} as well as LSST \citep[][]{Ivezic2008} employ all or nearly all of these bands.
Thus, a substantial and expanding fraction of the sky is covered by deep photometry in at least \gband{}\rband{}\iband{}\zband{}.

\begin{figure}
    \includegraphics[width=0.5\textwidth]{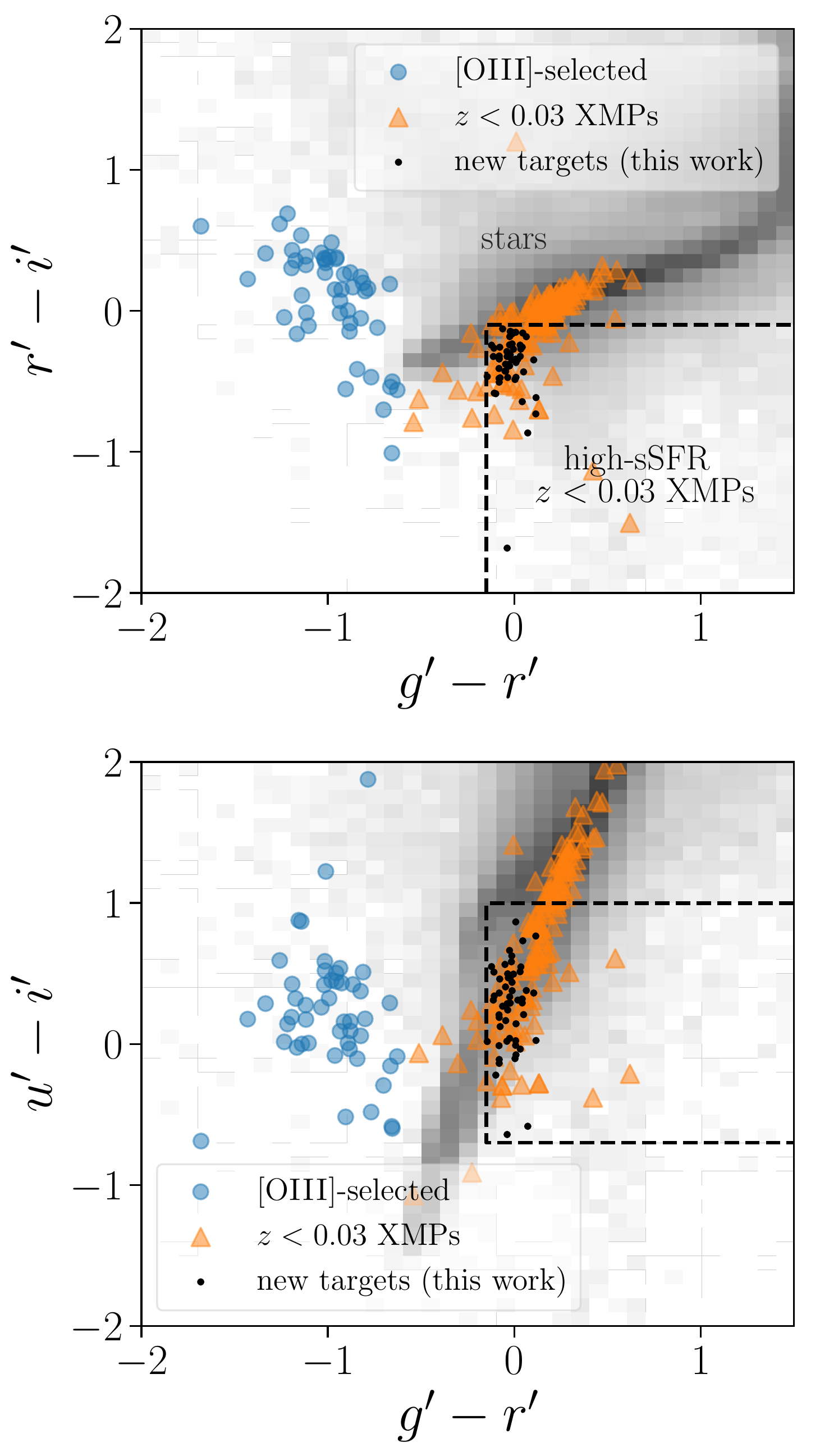}
    \caption{
        Color-color plots highlighting the position of our new targets alongside known systems in $\gband{}-\rband{}$, $\rband{}-\iband{}$, and $\uband{}-\iband{}$.
        We plot the position of the Blueberry galaxies \citep[$\rband{}-\iband{}$ selected Green Pea analogues,][; blue circles]{Yang2017} and known XMP galaxies drawn from the literature \citep[orange triangles:][]{SanchezAlmeida2016,Guseva2017}.
        The background density histogram represents spectroscopically-confirmed stars drawn from SDSS.
        The $z<0.03$ XMPs cluster near $\gband{}-\rband{}\simeq 0$ due to a similar excess in both \gband{} and \rband{} (Figure~\ref{fig:sedbands}).
        Based upon the distinctive colors of these spectroscopically-confirmed galaxies, we defined a set of color cuts (dashed black lines) designed to identify previously unknown high-sSFR XMPs using only five-band SDSS photometry.
        Projections of the color cuts and and the 53 target galaxies selected in this paper are displayed as black dashed lines and black points, respectively.
    }
    \label{fig:gri_prop}
\end{figure}

Star-forming dwarf galaxies have distinctive optical spectra, dominated by a blue continuum and the strong emission lines H$\alpha$, H$\beta$, and [\oiii{}] $\lambda \lambda 4959,5007$.
In systems with sufficiently dominant recent star formation, this nebular emission can heavily contaminate broadband photometry and produce distinctive color excesses.
In Figure~\ref{fig:gri_prop} we plot the $\gband{}-\rband{}$ versus $\rband{}-\iband{}$ and $\uband{}-\iband{}$ SDSS colors of spectroscopically-confirmed XMPs, alongside those of star-forming galaxies previously selected using color excess techniques and the color distribution for stars.
Since we are interested in seeking out the lowest-metallicity galaxies, and we expect these galaxies to be relatively low-mass and thus faint \citep[e.g.][]{Berg2012}, we focus on the nearest galaxies at the lowest redshifts $z<0.03$ (luminosity distances $\lesssim 100$ Mpc).
We plot $z<0.03$ XMPs uncovered from the SDSS spectroscopic dataset by \citet{SanchezAlmeida2016} and extremely metal-deficient galaxies catalogued by \citet{Guseva2017}, which collectively include both systems with very intense ongoing star formation as well as objects with low gas-phase oxygen abundances but relatively low equivalent-width emission.
For comparison, we plot the Blueberry galaxies presented by \citet{Yang2017}, which are $z<0.03$ systems selected on [\oiii{}]+H$\beta$ emission; these are low-redshift counterparts to the [\oiii{}]-selected Green Pea galaxies, which reside at $z\sim$ 0.1--0.3 \citep{Cardamone2009}.
Especially at the very faintest magnitudes, star--galaxy separation algorithms can confuse compact star-forming galaxies for point sources --- hence we also plot the distribution of spectroscopically-identified stars in this figure.
These plots reveal that extremely metal-deficient systems occupy a region of color-color space disjoint from that of galaxies selected photometrically via an [\oiii{}] excess at the same low redshifts, and overlap more significantly with the stellar locus.

\begin{figure}
    \includegraphics[width=0.5\textwidth]{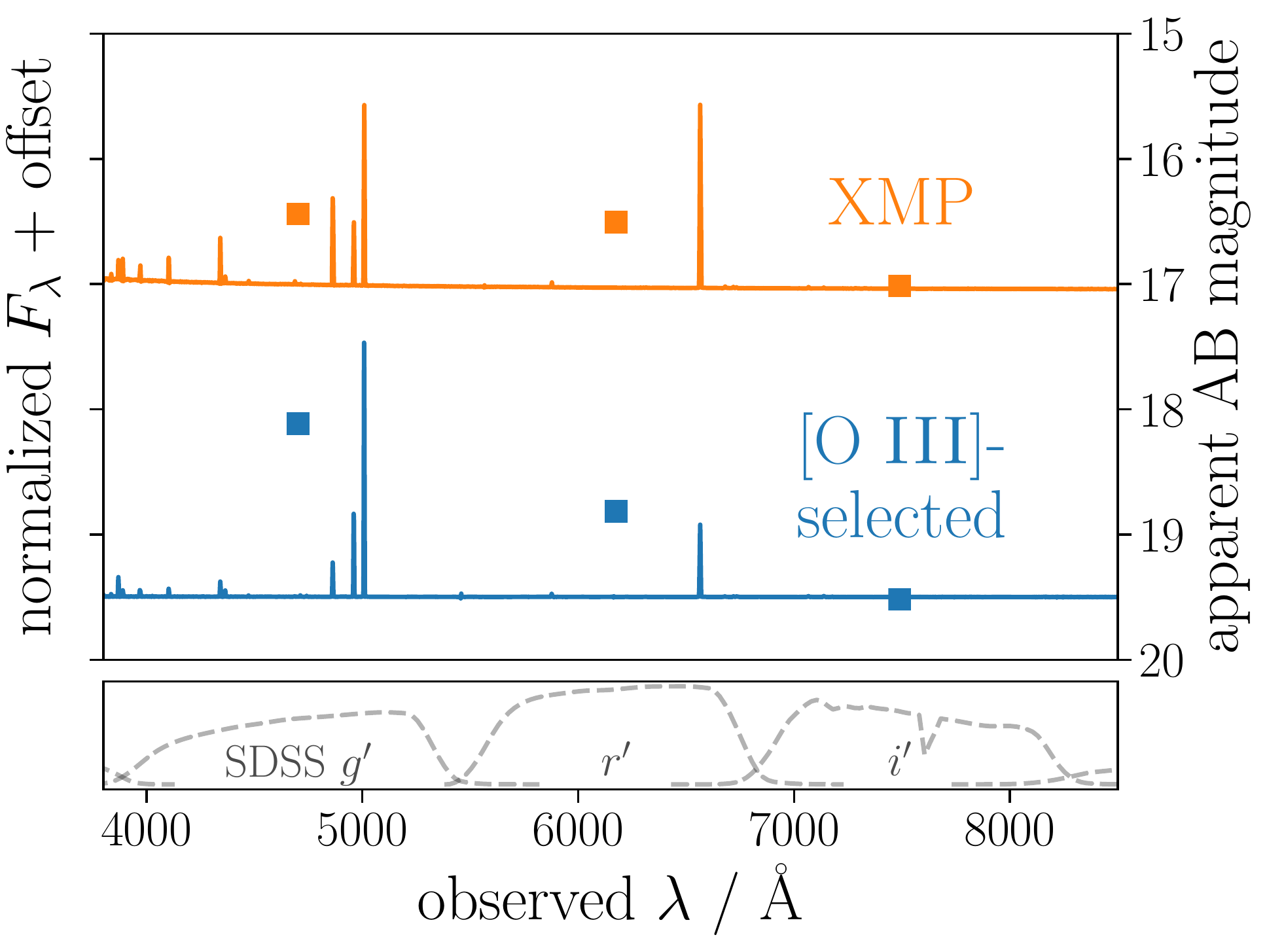}
    \caption{
        SEDs and SDSS photometry for two prototypical nearby ($z<0.03$) star-forming galaxies.
        The grayscale curves in the lower panel represent the throughput of the SDSS \gband{}, \rband{}, and \iband{} bands which we focus on in this work.
        The bottom blue SED and photometry is from a blueberry galaxy (RA, Dec 13:23:47.46, -01:32:52.00) and is typical of galaxies selected by a single band excess; strong [\oiii{}] emission produces an excess in \gband{} relative to both \rband{} and \iband{}.
        On the other hand, the top galaxy is the XMP I Zw 18 NW (RA, Dec 09:34:02.03, +55:14:27.79), and shows much lower [\oiii{}]/H$\alpha$ due to its low metallicity.
        In this case, [\oiii{}]+H$\beta$ and H$\alpha$ produce nearly identical excesses, yielding a $\gband{}-\rband{}\sim 0$.
        Our selection is designed to capture systems like I Zw 18 with very low metallicity and relatively weak [\oiii{}] emission.
    }
    \label{fig:sedbands}
\end{figure}

In order to understand this offset in color-color space, we must examine the spectral energy distribution of nearby star-forming galaxies in more detail.
In Figure~\ref{fig:sedbands}, we plot the SDSS spectra and photometry for I Zw 18 NW alongside that of a galaxy selected using the [\oiii{}]-excess photometric criteria presented by \citet{Yang2017}.
In the latter system, [\oiii{}] $\lambda 5007$ is significantly stronger than H$\alpha$.
The [\oiii{}] and H$\alpha$ lines occupy the \gband{} and \rband{} bands for objects such as these at redshifts $z<0.03$.
As visible in Figure~\ref{fig:sedbands}, systems with strong [\oiii{}] emission relative to H$\alpha$ will produce a substantially larger photometric excess in \gband{} (from [\oiii{}]+H$\beta$) than in \rband{} (H$\alpha$), forming the basis of the Green Pea-like Blueberry galaxy selection (Figure~\ref{fig:gri_prop}).
For a fixed star formation history and gas geometry, the ratio of [\oiii{}]/H$\alpha$ peaks at a metallicity near $12+\log\mathrm{O/H}=8$ and decreases at lower metallicities \citep[e.g.][]{Kewley2002}.
As a result, selections based upon high [\oiii{}]/H$\alpha$ will preferentially identify systems at moderately subsolar metallicities ($Z/Z_\odot \simeq 0.2$).
This is evident in the metallicity distribution of the Green Pea galaxies \citep{Amorin2010}.

The colors of most actively star-forming XMPs at $z<0.03$ are significantly different from these higher-metallicity systems.
Figure~\ref{fig:sedbands} reveals that the [\oiii{}] $\lambda 5007$ emission line presents approximately the same flux as H$\alpha$ in I Zw 18 NW.
As a result, the $\gband{}-\rband{}$ color of this object is nearly zero, and it would be missed by a blue $\gband{}-\rband{}$ selection designed to identify strong [\oiii{}] emitters (Figure~\ref{fig:gri_prop}).
However, both the $\gband{}-\iband{}$ and $\rband{}-\iband{}$ colors are significantly blue in this XMP due to contamination by [\oiii{}]+H$\beta$ and H$\alpha$ (respectively) relative to the continuum in \iband{}.
By identifying an excess in both \gband{} and \rband{} relative to \iband{} and focusing on $\gband{}-\rband{}$ colors near zero, we can define new color criteria to select star-forming XMPs.

This physical picture motivates development of a two-band excess selection (in \gband{} and \rband{}) for identifying XMPs at low redshift.
We define the color cuts used for this work by examining the measured colors of known XMPs ($12+\log\mathrm{O/H}<7.7$) at redshifts $z<0.03$ culled from the literature \citep[][and references therein; Figure~\ref{fig:gri_prop}]{Kunth2000,Shirazi2012,Guseva2017,Izotov2018}.
In particular, we focus on systems with large specific star formation rates and high equivalent width emission indicative of a substantial population of massive stars.
\begin{itemize}
    \item $\gband{}-\iband{} < -0.1 \;\; (5\sigma)$, $\rband{}-\iband{} < -0.1 \;\; (5\sigma)$: Statistically-significant excesses in both the \gband{} and \rband{} bands relative to \iband{}. For ideal tophat filters, an excess of 0.1 magnitudes over a flat continuum in $F_\nu$ corresponds to a nebular flux contribution at 10\% the total flux of the continuum in the bandpass.
    \item $-0.15 < \gband{}-\rband{}$: An \rband{}-band excess comparable to or stronger than that in \gband{} (see Figure~\ref{fig:sedbands}).
    \item $-0.7 < \uband{}-\iband{} < 1.0$: A relatively flat continuum in $\uband{}-\iband{}$ (large scatter, due to both real variations in star formation history and shallower $\uband{}$-band constraints).
    \item $-0.3 < \iband{}-\zband{} < 0.3$: A flat continuum in $\iband{}-\zband{}$.
\end{itemize}
We utilize the \texttt{modelMag} measurements included in the SDSS database which are derived from the better-fitting of a de Vaucouleurs and an exponential profile model.
These measurements are thus appropriate for both point-like and extended objects.

Single-band excess selection techniques such as those defining the Green Peas have the advantage of isolating galaxies with particularly unique colors.
As hinted at by the overlap with the stellar locus in Figure~\ref{fig:gri_prop}, our color selection includes some contaminants which are dominated by stars of A and F type and a smaller number of quasars (predominantly at $z\sim2.3$).
For the purposes of this initial study, we utilize morphology cuts to refine our sample.
We require that our targets are classified as non-pointlike (extended) by the SDSS algorithm,
In addition, we crossmatched point-like SDSS candidates with the deeper Hyper Suprime-Cam Subaru Strategic Program \citep[HSC-SSP;][]{Aihara2018} DR1 photometry and identified a small number of unresolved SDSS objects categorized as extended in this catalog, which we included in our target list.
Of the set of 959 objects in SDSS DR13 which satisfy our photometric selection, 148 have associated SDSS spectra.
All are emission line galaxies at low redshift $z<0.08$ (all but five reside at $z<0.03$), with median H$\beta$ equivalent width 96 \AA{} indicative of substantial recent star formation.
By design, this observed subset includes many systems in the XMP regime, including canonical systems such as I Zw 18.

Motivated by the successful test of this selection, we initiated a spectroscopic campaign to target the fainter objects selected by our cuts which lack SDSS spectra.
We manually cleaned the set of targets without SDSS spectra to identify high-quality candidates for spectroscopic follow-up.
In particular, we removed giant \hii{} regions embedded in spiral galaxies or in general any larger system with an existing SDSS spectrum.
We then crossmatched with SIMBAD \citep{Wenger2000} and excised any object with a metallicity measurement in the literature.
We also discarded any obvious photometric artifacts.
The majority of candidates removed by this cleaning are discarded as \hii{} components of substantially larger and well-studied star-forming galaxies.
From the 959 extended objects in the SDSS DR13 photometric database satisfying our selection criteria, this procedure yielded 161 high-quality cleaned targets awaiting spectroscopic confirmation.

\begin{figure}
    \includegraphics[width=0.5\textwidth]{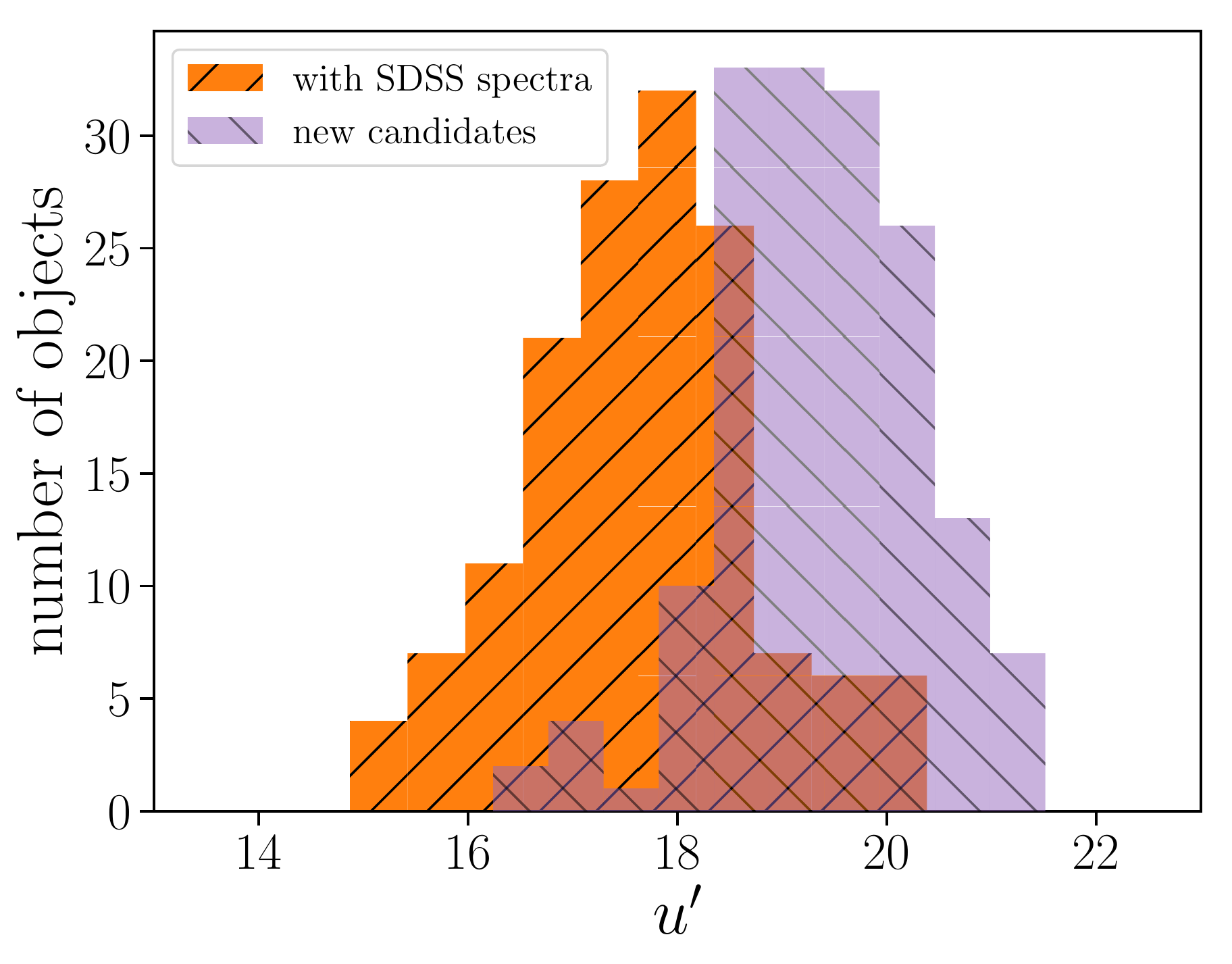}
    \caption{
        A histogram representing the \uband{}-band magnitude distribution of objects satisfying our photometric selection criteria, including both the 148 objects with flux-matched SDSS spectra and the 161 new cleaned candidates identified for follow-up.
        The subset with SDSS spectra drops off steeply at magnitudes fainter than $\uband{}>18$ due to the apparent magnitude limit of this survey.
        Our new targets peak at a significantly fainter magnitude $\uband{}=19$, and extend as faint as $\uband{}=21$, significantly below the completeness limits of the SDSS spectra.
    }
    \label{fig:magdist_select}
\end{figure}

We plot the \uband{}-band apparent magnitude distribution of cleaned targets satisfying our photometric selection in Figure~\ref{fig:magdist_select}.
The objects that were observed spectroscopically by SDSS have a median \uband{}-band magnitude of 17.7, and extend down only to 20.4.
The turnover near $\uband{}=18$ is consistent with the design of the SDSS spectroscopic survey, which is complete to galaxies brighter than $\rband{}<18$ \citep{Strauss2002}.
In contrast, the distribution of the cleaned photometric targets selected for study in this work peaks at $\uband{}\sim 19$ and extends down to nearly $\uband{}=22$.
Our sample is likely significantly incomplete at $\uband{}\gtrsim$ 19--20 due to the increasing uncertainty of the SDSS star--galaxy separation algorithm and color measurements involving the shallower $\iband{}$ and $\zband{}$ bands at these faint magnitudes.
Even so, our photometric selection probes significantly below the completeness limits of the SDSS spectroscopic survey.
By locating fainter galaxies with a band contamination technique trained at redshifts $z<0.03$ (see above), we aim to access lower-mass galaxies and potentially a higher fraction of XMPs.
Deeper photometric surveys such as LSST will likely dramatically expand the population of objects accessible to this technique.

\begin{table*}
    \centering
    \caption{Coordinates, SDSS photometry, and exposure times for the 53 objects we obtained [\oiii{}] $\lambda 4363$ detections in (SNR$>2$) with MMT Red Channel or Blue Channel (RC or BC, respectively).
    Effective radii from exponential fits (accounting for the seeing point spread function) in the \iband{}-band reported by the SDSS photometric pipeline are also presented.
    The redshifts and distances inferred from the MMT spectra along with absolute SDSS \texttt{modelMag} \iband{} measurements corrected accordingly are presented in the final columns (Section~\ref{sec:phot}).
    The confirmed galaxy for which we determined extendedness using the HSC-SSP catalog rather than SDSS photometry (SDSS J0845+0131) is denoted by a $\ast$.
    The IAU designations for each SDSS photometric object are presented here, though shortened names are used in subsequent tables and the text.
    }
    \label{tab:targs}
\begin{tabular}{lccccccccccc}
\hline
Name & \uband{} & \gband{} & \rband{} & \iband{} & \zband{} & $r_{\mathrm{eff}}(i)$ & instrument & exposure & $z$ & distance & $M_i$\\ 
 &  &  &  &  &  & arcsec (SDSS) & (MMT) & time (s) &  & Mpc & \\ 
\hline
SDSS J003740.45+291010.0 & 19.38 & 18.73 & 18.77 & 19.10 & 18.88 & 3.4 & RC & 1200 & 0.017 & 77 & -15.33\\ 
SDSS J005344.44-093725.9 & 19.08 & 18.33 & 18.44 & 18.77 & 18.87 & 3.1 & RC & 1200 & 0.014 & 69 & -15.42\\ 
SDSS J011519.73+270907.7 & 19.72 & 19.09 & 19.23 & 19.70 & 19.65 & 0.5 & RC & 1200 & 0.013 & 59 & -14.15\\ 
SDSS J015351.12+261548.9 & 18.04 & 17.24 & 17.26 & 17.57 & 17.41 & 3.9 & RC & 1200 & 0.010 & 47 & -15.80\\ 
SDSS J015943.87-062232.9 & 18.65 & 17.90 & 17.94 & 18.41 & 18.33 & 2.8 & RC & 900 & 0.0093 & 47 & -14.93\\ 
SDSS J020321.99+220402.1 & 19.99 & 19.54 & 19.62 & 20.13 & 19.98 & 1.4 & RC & 1500 & 0.0093 & 44 & -13.08\\ 
SDSS J020540.72-173252.4 & 17.26 & 16.41 & 16.37 & 16.52 & 16.45 & 3.0 & RC & 900 & 0.016 & 76 & -17.88\\ 
SDSS J020601.09+271211.4 & 19.30 & 18.38 & 18.50 & 18.75 & 18.66 & 3.0 & RC & 900 & 0.017 & 76 & -15.64\\ 
SDSS J020613.35-060948.8 & 19.04 & 18.44 & 18.52 & 18.86 & 18.90 & 0.8 & RC & 900 & 0.014 & 64 & -15.18\\ 
SDSS J022225.97-035509.9 & 19.07 & 18.17 & 18.13 & 18.77 & 18.99 & 1.0 & RC & 900 & 0.0084 & 42 & -14.34\\ 
SDSS J024441.11+284600.7 & 19.75 & 18.71 & 18.76 & 19.19 & 19.08 & 2.1 & RC/BC & 3600 & 0.020 & 88 & -15.54\\ 
SDSS J024758.00+035025.0 & 19.65 & 19.44 & 19.37 & 20.24 & 20.16 & 0.4 & RC & 1200 & 0.0041 & 22 & -11.51\\ 
SDSS J030024.01+020628.8 & 19.55 & 19.26 & 19.34 & 19.66 & 19.61 & 1.0 & RC & 1200 & 0.021 & 92 & -15.15\\ 
SDSS J031911.82+401729.7 & 20.15 & 19.51 & 19.56 & 19.99 & 19.81 & 0.9 & RC & 1500 & 0.019 & 80 & -14.54\\ 
SDSS J043015.89+084530.5 & 19.34 & 18.31 & 18.41 & 19.00 & 18.96 & 1.3 & BC & 1500 & 0.012 & 49 & -14.44\\ 
SDSS J044412.94-045819.9 & 18.35 & 17.47 & 17.58 & 17.84 & 17.73 & 1.6 & BC & 1800 & 0.016 & 68 & -16.31\\ 
SDSS J051134.12-003818.8 & 19.25 & 18.62 & 18.60 & 18.94 & 18.93 & 1.7 & RC & 1200 & 0.015 & 64 & -15.07\\ 
SDSS J061046.04+644950.1 & 18.63 & 18.18 & 18.20 & 18.59 & 18.55 & 0.8 & BC & 1800 & 0.016 & 67 & -15.53\\ 
SDSS J064101.26+381532.9 & 20.55 & 19.88 & 19.95 & 20.43 & 20.15 & 0.8 & RC & 1800 & 0.024 & 100 & -14.57\\ 
SDSS J075733.93+475030.7 & 19.09 & 18.50 & 18.45 & 18.89 & 18.79 & 2.9 & RC & 1800 & 0.0076 & 31 & -13.58\\ 
SDSS J081713.08+263350.3 & 20.21 & 19.44 & 19.44 & 19.71 & 19.55 & 1.3 & RC & 1500 & 0.019 & 80 & -14.79\\ 
SDSS J082225.73+460522.8 & 20.49 & 19.70 & 19.78 & 20.13 & 19.91 & 1.3 & RC/BC & 4200 & 0.013 & 52 & -13.46\\ 
SDSS J083609.47+030600.0 & 21.32 & 19.94 & 19.83 & 20.56 & 20.82 & 1.1 & BC & 2700 & 0.0094 & 38 & -12.35\\ 
SDSS J084425.44-024001.1 & 19.72 & 19.22 & 19.21 & 19.58 & 19.48 & 3.7 & RC & 900 & 0.010 & 42 & -13.53\\ 
SDSS J084530.80+013151.2 $\ast$ & 22.17 & 21.09 & 21.13 & 22.81 & 21.91 & 3.4 & RC & 1800 & 0.013 & 53 & -10.80\\ 
SDSS J090541.82+253227.9 & 18.86 & 18.07 & 18.04 & 18.31 & 18.18 & 3.4 & RC & 2100 & 0.0093 & 38 & -14.62\\ 
SDSS J090700.00-000251.2 & 19.10 & 18.17 & 18.20 & 18.60 & 18.38 & 1.6 & RC & 1200 & 0.019 & 79 & -15.89\\ 
SDSS J091737.69+505715.6 & 20.23 & 19.51 & 19.48 & 19.72 & 19.96 & 1.0 & RC/BC & 3300 & 0.012 & 51 & -13.82\\ 
SDSS J092713.61+031424.6 & 19.76 & 19.24 & 19.12 & 19.74 & 19.77 & 0.9 & RC & 1500 & 0.0041 & 15 & -11.18\\ 
SDSS J093751.74+204653.2 & 21.07 & 20.61 & 20.71 & 21.29 & 21.31 & 0.5 & RC & 1500 & 0.020 & 83 & -13.29\\ 
SDSS J095506.70-031617.7 & 18.55 & 17.80 & 17.82 & 17.97 & 17.93 & 3.0 & RC & 900 & 0.018 & 77 & -16.46\\ 
SDSS J100008.52+273614.9 & 20.32 & 19.99 & 19.99 & 20.33 & 20.34 & 0.5 & RC & 2100 & 0.014 & 62 & -13.63\\ 
SDSS J100438.64+291146.2 & 18.77 & 18.31 & 18.33 & 18.48 & 18.52 & 1.2 & RC & 1500 & 0.015 & 66 & -15.60\\ 
SDSS J100512.15+372201.5 & 19.34 & 19.09 & 19.06 & 19.38 & 19.52 & 2.3 & RC & 15300 & 0.0014 & 2.7 & -7.77\\ 
SDSS J101818.56+213736.6 & 20.49 & 19.87 & 19.88 & 20.11 & 20.22 & 0.7 & RC & 1800 & 0.020 & 87 & -14.60\\ 
SDSS J102500.91+572129.9 & 18.61 & 18.26 & 18.30 & 18.59 & 18.57 & 1.2 & RC/BC & 4800 & 0.0078 & 35 & -14.15\\ 
SDSS J103129.68+383210.9 & 18.74 & 18.12 & 18.13 & 18.30 & 18.26 & 1.8 & RC/BC & 4200 & 0.011 & 47 & -15.05\\ 
SDSS J103207.91+562129.6 & 19.21 & 18.28 & 18.30 & 18.54 & 18.80 & 2.5 & RC/BC & 4200 & 0.024 & 104 & -16.53\\ 
SDSS J103400.94-022158.9 & 16.24 & 15.74 & 15.68 & 15.86 & 15.88 & 8.9 & RC & 1200 & 0.0058 & 25 & -16.15\\ 
SDSS J103641.07+140715.1 & 19.98 & 19.37 & 19.27 & 19.61 & 19.72 & 1.1 & RC & 2100 & 0.0084 & 38 & -13.31\\ 
SDSS J110431.76-004146.5 & 18.57 & 17.77 & 17.85 & 18.11 & 18.03 & 2.9 & RC & 1200 & 0.019 & 82 & -16.46\\ 
SDSS J110521.68+175417.8 & 18.56 & 18.02 & 18.04 & 18.22 & 18.19 & 1.7 & RC & 1200 & 0.010 & 47 & -15.12\\ 
SDSS J110849.73+145358.5 & 18.93 & 18.56 & 18.55 & 19.04 & 18.98 & 1.9 & RC & 1200 & 0.012 & 54 & -14.62\\ 
SDSS J111857.75+670953.3 & 19.61 & 18.60 & 18.59 & 18.75 & 18.47 & 1.2 & RC/BC & 4800 & 0.0089 & 41 & -14.29\\ 
SDSS J112712.36+391520.9 & 18.95 & 18.58 & 18.57 & 18.92 & 18.77 & 4.5 & RC & 1500 & 0.0067 & 33 & -13.64\\ 
SDSS J113058.67+234836.9 & 18.23 & 17.68 & 17.76 & 18.24 & 18.16 & 1.9 & RC & 900 & 0.0067 & 33 & -14.36\\ 
SDSS J120459.62+525934.6 & 19.53 & 19.00 & 18.96 & 19.21 & 19.32 & 1.6 & RC & 1500 & 0.012 & 54 & -14.45\\ 
SDSS J120652.26+173942.5 & 20.05 & 19.07 & 19.08 & 19.43 & 19.21 & 1.3 & RC & 1200 & 0.013 & 59 & -14.42\\ 
SDSS J122641.86+223034.2 & 20.21 & 19.82 & 19.81 & 20.28 & 20.15 & 1.0 & RC & 1200 & 0.018 & 79 & -14.21\\ 
SDSS J123034.76+370821.3 & 19.17 & 18.71 & 18.77 & 18.90 & 18.82 & 1.0 & RC & 1200 & 0.022 & 96 & -16.01\\ 
SDSS J123920.18+392104.5 & 19.10 & 18.41 & 18.49 & 18.90 & 18.86 & 3.0 & RC & 1200 & 0.019 & 82 & -15.68\\ 
SDSS J125134.65+392642.6 & 20.36 & 19.48 & 19.51 & 19.89 & 19.70 & 0.8 & RC & 1200 & 0.0076 & 37 & -12.98\\ 
SDSS J125727.52+270605.3 & 20.92 & 20.09 & 20.14 & 20.51 & 20.26 & 0.8 & RC & 1200 & 0.019 & 83 & -14.07\\ 
\hline
\end{tabular}
\end{table*}

\begin{table}
    \centering
    \caption{Summary of the MMT nights allocated for this program.
    Seeing reported was from wavefront sensor measurements (typically 2--3 performed per night).}
    \label{tab:obs}
\begin{tabular}{lccl}
\hline
    Date & Instrument & Seeing \\
\hline
    11/1/18 & MMT Red Channel (RC) & 1.0$''$  \\
    12/1/18 & RC & 0.7--2$''$  \\
    13/1/18 & RC & 0.6$''$  \\
    14/1/18 & RC & 1--2$''$ \\
\hline
    22/1/18 & MMT Blue Channel (BC) & 0.7$''$  \\
    23/1/18 & BC & 1.5--2$''$ \\
    24/1/18 & BC & 2$''$  \\
\hline
\end{tabular}
\end{table}

\begin{figure*}
    \includegraphics[width=0.8\textwidth]{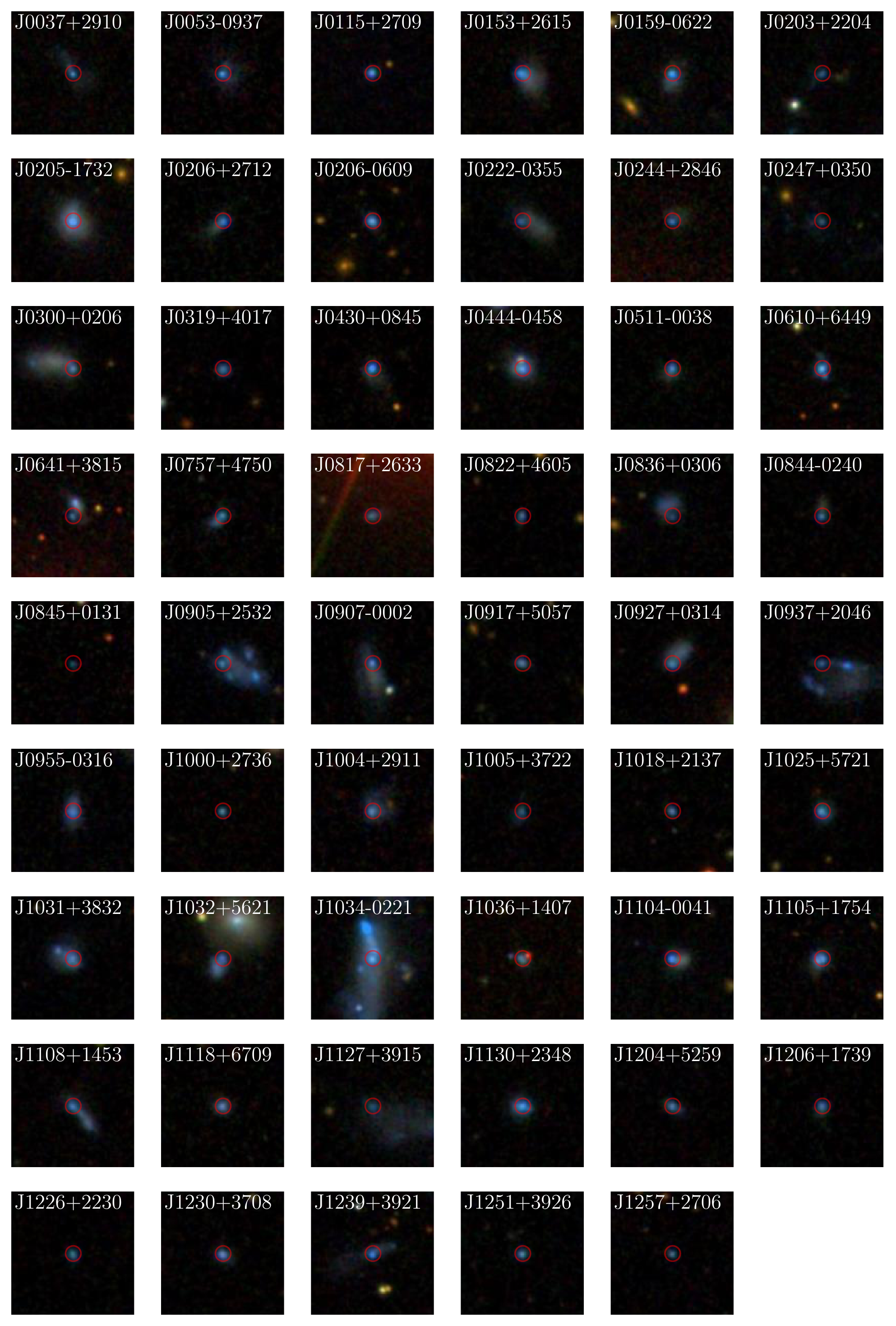}
    \caption{SDSS \gband{}\rband{}\iband{} mosaic images centered on our spectroscopic targets.
    A 3$''$ radius red circle is centered on the target region observed; this radius corresponds to transverse comoving distances ranging from 0.04--1.5 kpc at the distances of these objects.
    The majority are dominated in the optical by the compact star-forming regions targeted in each, for which we obtained MMT spectroscopic follow-up.
    }
    \label{fig:cutouts}
\end{figure*}

\section{Observations and Analysis}
\label{sec:obsanalysis}

With a sample of new star-forming XMP candidates in-hand, we pursued spectroscopic observations to confirm and characterize these objects.
Over the course of seven nights with the MMT in January 2018, we detected the auroral line [\oiii{}] $\lambda 4363$ \AA{} and measured direct-temperature metallicities in 53 systems selected from our sample on the basis of visibility --- their photometric properties are summarized in Table~\ref{tab:targs}.
The range of apparent magnitudes of the observed targets presented here is representative of that of the full set of 161 SDSS photometric targets, with an \iband{}-band $16^{\mathrm{th}}$--$50^{\mathrm{th}}$--$84^{\mathrm{th}}$ percentile range of 18.3--19.0--20.1 compared to 18.2--19.0--20.2 for the full cleaned sample.
In the following subsections, we describe the photometric data and bulk galaxy constraints derived from SED modeling; then discuss the spectroscopic observations and data reduction; and describe the measurement of nebular emission lines and metallicities from these spectra.

\subsection{Photometric data and distances}
\label{sec:phot}

The SDSS imaging of our targets provides constraints on the magnitude of their recent star formation.
We plot SDSS mosaic images of the 53 targets presented in this paper in Figure~\ref{fig:cutouts}.
The majority of galaxies targeted are dominated by the compact region we have selected for follow-up, so that the SDSS \texttt{modelMag} photometry of our targets represents essentially the entire galaxy in the optical; though several systems appear to be part of a larger irregular structure of embedded star-forming regions.
Since our spectroscopic data is limited to the compact region covered by our slit we will focus our attention on these regions, while also considering the impact of larger aperture photometry.

We fit the SDSS \texttt{modelMag} flux measurements with \texttt{Prospector}, a \texttt{python} framework for stellar population synthesis modeling \citep{Johnson2017,Leja2017}\footnote{\url{https://github.com/bd-j/prospector}}.
In particular, we utilize the affine-invariant MCMC sampler \texttt{emcee} \citep{Foreman-Mackey2013} and the Flexible Stellar Population Synthesis (FSPS) models including nebular emission, described by \citet{Conroy2009,Conroy2010} and \citet{Byler2017}.
We adopt a constant star formation history representing the most recent activity in these systems, which we find to be an adequate fit to the SDSS \uband{}\gband{}\rband{}\iband{}\zband{} photometry.
Including the nebular emission model implemented by \citet{Byler2017} allows us to model the entire SED including contaminated bands.
We adopt a \citet{Chabrier2003} IMF with the Padova isochrones and MILES spectral library.
For each primary adjustable parameter, we established a flat prior likelihood distribution and ensured visually that no fits were forced to converge at the edges of this parameter space.
In particular, we allow metallicity to vary over the range $-2 < \log_{10}(Z/Z_\odot) < 0.2$; dust extinction to vary up to an optical depth of 0.5; and the gas ionization parameter $\log U$ to vary over the full range of the models ($-4<\log U<0$).
Since our broadband optical photometry and optical spectra are insufficient to directly constrain the stellar metallicity, we adopt the standard assumption that the stellar and gas-phase metallicities are matched with solar abundances for the purposes of this work.
Stellar mass estimates and effective ages are extracted by marginalizing over the full posterior distribution and presented in Table~\ref{tab:linemet}.

Very young $\lesssim$ 10 Myr stellar populations and the nebular emission they excite likely dominate the UV--optical spectra of assembling galaxies at the sSFRs inferred in the reionization era \citep[e.g.][]{Faisst2016}.
We have designed our search technique to locate XMPs with similarly high sSFR in order to study comparable systems at low metallicity nearby.
In the fitting procedure above we have adopted a constant star formation history, which minimizes the number additional of free parameters to two (effectively, the scale and duration of star formation).
This provides a good fit to the optical photometry including the bands probing clean continuum (\uband{}, \iband{}, and \zband{}) as well as the nebular-contaminated \gband{} and \rband{} bands.
Importantly, this model is also commonly-used to derive star formation rates and stellar masses for galaxies in the reionization era \citep[e.g.][]{Labbe2013,Stark2013,Salmon2015}, enabling more direct comparison with results at these redshifts.

Two important systematic uncertainties may impact the results of our photometric SED fitting.
First, we assess the degree to which the SDSS \texttt{modelMag} measurements capture the total flux of the galaxies under study.
To estimate an upper limit to the total brightness of our systems, we identify the full extent of the galaxies in the \gband{}-band by producing a segmentation map with the \texttt{photutils} software package \citep{Bradley2018} with a limiting SNR of 1.5 and expanding it by 10 pixels in all directions, which we verify visually encompasses all diffuse extended light visible in the images.
We then measure a total magnitude in each band by summing flux within this full galaxy footprint.
The ratio of this total flux in the \iband{}-band to that represented by the \texttt{modelMag} measurements is typically only 1--3 (with 16--50--84$^{\mathrm{th}}$ percentile confidence interval $1.4^{+1.9}_{-0.4}$), with only four systems yielding total flux estimates 10--40 times larger than the \texttt{modelMag}.
This indicates that the compact burst we are focusing on typically represents a significant fraction of the entire galaxy in the optical.

Second, we consider the impact of changing our assumed star formation history.
The optical photometry of many of the objects in our sample require very young ages ($<20$ Myr; Table~\ref{tab:linemet}) to reproduce, indicating that the systems are undergoing a burst or significant recent upturn in star formation rate.
The young, massive stars formed in this burst will outshine older stars from earlier episodes of star formation in both the UV and optical, making it difficult to reliably constrain the total stellar mass.
Resolved stellar population studies working with very nearby galaxies with similar properties like I Zw 18 indicate that these older stars can contribute significantly to the total stellar mass of such galaxies \citep[e.g.][]{Aloisi2007,Corbin2008}.
While we cannot directly constrain the star formation history of our targets with unresolved optical photometry, we can assess the systematic uncertainty introduced by our assumed star formation history.
To this end, we also estimate stellar masses utilizing the \uband{},\iband{} mass-to-light ratio calibration derived by \citet{Bell2003} assuming a more extended exponential star formation history.
The total stellar masses computed with this calibration corrected to a \citep{Chabrier2003} IMF and using our total \uband{},\iband{} magnitudes found above are also provided in Table~\ref{tab:linemet}.
These total stellar masses are $21^{+46}_{-9}$ times larger than those found for the recent burst (from our constant star formation history fits), primarily due to the the implicit prior on a substantially more extended star formation history.
We proceed with the mass and age estimates computed with \texttt{Prospector} assuming a constant star formation history for consistency with studies at $z>6$, but caution that these measurements likely underestimate the total stellar mass formed in these systems over the age of the universe.

For distances, we rely on redshifts measured from fitting the strong emission lines in the MMT spectra (see Section~\ref{sec:specobs}).
Since several galaxies reside at very low redshifts where the Hubble flow does not yet dominate over peculiar velocities, we utilize the \citet{Tonry2000} local flow model with $H_0=70 \unit{km s^{-1} Mpc^{-1}}$ to convert redshifts to distances.
The adoption of this local flow model is necessary as no other distance indicators are readily available; while it inevitably introduces systematic uncertainties in derived measurements at the lowest redshifts, only the absolute magnitudes and masses presented in this paper are affected.
We note that this uncertainty is most significant for J1005+3722, which is the only galaxy with inferred distance $<15$ Mpc (at $z=0.0014$, corresponding to 2.7 Mpc in our model).
The measured redshifts and adopted distances are presented in Table~\ref{tab:targs}.

\subsection{Spectroscopic observations}
\label{sec:specobs}

We obtained seven nights on the 6.5m MMT in 2018A to follow-up our XMP candidates.
Because of a dewar issue with the Blue Channel spectrograph, we used the Red Channel spectrograph for the first four nights of our program (January 11--14 2018, Table~\ref{tab:obs}).
We used the 300 $\mathrm{lines}/\mathrm{mm}$ grating with the $1.5''\times 180''$ slit, providing simultaneous coverage over 3500--6650 \AA{} (from [\oii{}] $\lambda \lambda 3727,3729$ to H$\alpha$) with spectral resolution 11 \AA{} FWHM (measured from nebular emission lines).
The slit was aligned at parallactic to minimize losses on the blue end.
Exposure times were adjusted to obtain by-eye significant detections of the [\oiii{}] $\lambda 4363$ auroral line for direct electron temperature $T_e$ metallicity measurement, and ranged from 15--30 minutes typically (see Table~\ref{tab:targs}).
Wavefront sensor corrections were performed at least twice per night, and indicated that seeing varied from 0.6--2$''$ over the course of the four nights (Table~\ref{tab:obs}).

Blue Channel was restored to the MMT in time for our three night block on January 22--24 2018 (Table~\ref{tab:obs}).
Our Blue Channel observations were similar: we used the 300 $\mathrm{lines}/\mathrm{mm}$ grating and $1.5''\times180''$ slit oriented at parallactic, which in this case yielded spectra covering 3100--8200 \AA{} with an effective spectral resolution of 7 \AA{} FWHM.
Winds and the resulting seeing were again variable, with the wavefront sensor reporting values ranging from 0.7--2$''$; all but one half night yielded usable data.

The spectra were reduced using standard longslit techniques.
Our calibrations consisted of three flux standard observations per night (of stars LB 227 and Feige 34) with HeAr/Ne lamp exposures taken at each to fix the wavelength solution and flats taken at the beginning of each night.
The data were reduced using a custom \texttt{python} reduction pipeline which found arc solutions and performed overscan bias correction and flat-fielding, median-combination of the individual data frames, rectification, background subtraction, and airmass-corrected flux normalization using the standard star observations.
The one-dimensional spectra were extracted with a boxcar of width $7''$ or $14''$, adjusted to ensure all object flux was captured while maximizing signal-to-noise.
The final one-dimensional spectra are uniformly dominated by a blue stellar continuum and strong nebular emission lines, with median signal-to-noise (S/N) of 14 per pixel in the continuum at 4550 \AA{} and median S/N of 25 in the total flux of H$\beta$.

\subsection{Nebular lines and metallicity}
\label{sec:nebmet}

\begin{figure*}
    \includegraphics[width=1.0\textwidth]{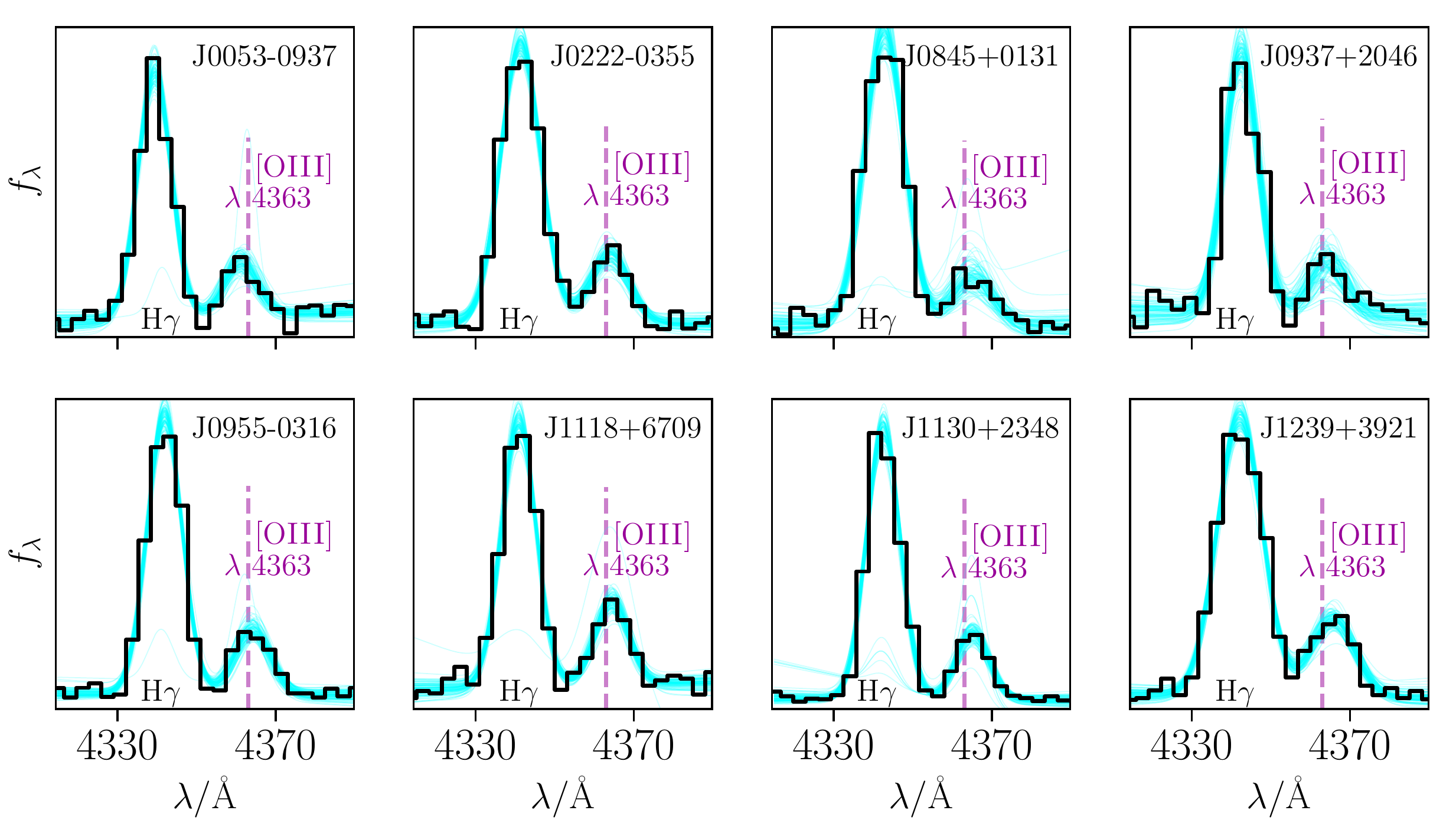}
    \caption{
        Detections and fits to the [\oiii{}] $\lambda 4363$ line for a subset of our targets.
        This line is critical for measurement of the electron temperature in the [\oiii{}]-emitting gas and thus for direct metallicity measurement.
        We fit this line simultaneously with the H$\gamma$ line adjacent to it to ensure accurate fluxes.
        The data is displayed in black, and sample models from the fit posterior are displayed in light blue.
    }
    \label{fig:oiii4363_fit}
\end{figure*}

Nebular lines were measured using the same software described in \citet{Senchyna2017}.
In summary, we conduct MCMC model fits consisting of Gaussians over a local linear continuum model using the \texttt{emcee} sampler \citep{Foreman-Mackey2013}.
This framework allows us to fit nearby lines jointly and obtain robust flux measurements.
A selection of the [\oiii{}] $\lambda 4363$ detections are displayed in Fig.~\ref{fig:oiii4363_fit}, demonstrating the typical S/N range achieved in this $T_e$-sensitive line.
All line fluxes were corrected for extinction assuming a Case B recombination ratio H$\alpha$/H$\beta$ of 2.74 \citep[$T_e=2\times 10^4 \unit{K}$, $n_e=10^3 \unit{cm^{-3}}$;][]{Draine2011}; if H$\alpha$/H$\beta$ was found to be less than this value (which occurred in only 7 systems, with a median offset from this theoretical value of $0.4 \sigma$), we assume negligible extinction and do not apply any corrections.
We adopt the \citet{Fitzpatrick1999} extinction curve, which differs minimally from the \citet{Gordon2003} SMC curve in the optical.
We found that varying the assumed intrinsic Case B value from 2.74--2.86 had minimal impact on the derived metallicities ($\lesssim 0.01$ dex).

With [\oiii{}] $\lambda 4363$ detections in hand, we measure gas-phase metallicities using the direct-$T_e$ method.
We adopt a two-zone photoionization model representing the [\oiii{}] and [\oii{}] emitting regions.
We obtain $T_e(\oiii{})$ measurements from the [\oiii{}] $\lambda 4363$ / [\oiii{}] $\lambda 5007$ flux ratios using the iterative method of \citet{Izotov2006}.
Since we lack confident [\oii{}] $\lambda\lambda 7320,7330$ detections for most of our objects, we estimate $T_e(\oii{})$ from the model-derived translation equations presented by \citet{Izotov2006}.
We then use the corrected Balmer, [\oiii{}] $\lambda\lambda 4959,5007$, and [\oii{}] $\lambda \lambda 3727,3729$ fluxes to derive $\mathrm{O/H}$ for each object.
The results are presented in Table~\ref{tab:linemet} alongside the stellar masses derived as described in Section~\ref{sec:phot}.
We also include measurements of several emission line equivalent widths and dust-corrected \rtt{} ($=$ [\oiii{}] $\lambda 4959$ + $\lambda 5007$ + [\oii{}] $\lambda \lambda 3727, 3729$ / H$\beta$) and \ott{} ($=$ [\oiii{}] $\lambda 4959$ + $\lambda 5007$ / [\oii{}] $\lambda \lambda 3727, 3729$) ratios.

In addition to the newly-observed objects whose MMT spectra we have presented above, we also consider the SDSS spectra of objects which fall within our photometric selection criteria and were observed as part of the SDSS survey.
Of the 148 of these SDSS spectra (see Section~\ref{sec:photsamp}), only 15 have $>2\sigma$ detections of both [\oii{}] $\lambda 3727,3729$ and [\oiii{}] $\lambda 4363$, largely due to the blue end limit of the original SDSS spectrograph.
In cases where the [\oii{}] $\lambda\lambda 3727,3729$ doublet is inaccessible, we utilize the \citet{Izotov2006} calibrations based on the weaker [\oii{}] $\lambda \lambda 7320,7330$ doublet.
In systems in which both [\oii{}] doublets are detected, we measure a median offset of 0.02 dex and $1\sigma$ scatter of 0.14 dex in the difference $\log\mathrm{O/H}_{3727}-\log\mathrm{O/H}_{7325}$.
This implies that the SDSS metallicity measurements are robust, but that we might expect greater scatter among the subset of systems lacking [\oii{}] $\lambda \lambda 3727,3729$ detections.

\begin{table*}
    \centering
    \caption{Basic spectroscopic and derived photometric properties for our sample.}
    \label{tab:linemet}
\begin{tabular}{lcccccccc}
\hline
Name & $W_0(\mathrm{H\beta})$ & $E(B-V)$ & $O_{32}$ & $R_{23}$ & $12+\log\mathrm{O/H}$ & age (Myr) & $\log_{10} (M_\star/M_\odot)$ & $\log_{10} [M_{\mathrm{tot}}($u',i'$)/M_\odot]$\\ 
SDSS & \AA{} &  &  &  &  & \texttt{prospector} & \texttt{prospector} & \citet{Bell2003}\\ 
\hline
J003740.45+291010.0& $159\pm31$& $0.14$& $4.7\pm0.3$& $8.0\pm0.6$& $7.78 \pm 0.19$& $25 \pm 13$& $6.59 \pm 0.17$& $8.13$\\ 
J005344.44-093725.9& $140\pm17$& $0.09$& $7.1\pm0.3$& $7.9\pm0.3$& $7.89 \pm 0.13$& $21 \pm 8$& $6.59 \pm 0.13$& $8.06$\\ 
J011519.73+270907.7& $131\pm10$& $0.31$& $7.5\pm0.3$& $7.0\pm0.3$& $7.56 \pm 0.12$& $16 \pm 3$& $5.98 \pm 0.07$& $7.63$\\ 
J015351.12+261548.9& $99\pm10$& $0.13$& $7.3\pm0.5$& $8.3\pm0.6$& $7.74 \pm 0.08$& $31 \pm 17$& $6.86 \pm 0.32$& $8.09$\\ 
J015943.87-062232.9& $148\pm17$& $0.10$& $8.5\pm0.6$& $7.1\pm0.4$& $7.55 \pm 0.12$& $13 \pm 3$& $6.34 \pm 0.14$& $7.85$\\ 
J020321.99+220402.1& $110\pm15$& $0.08$& $3.0\pm0.2$& $6.7\pm0.4$& $7.58 \pm 0.10$& $27 \pm 14$& $5.66 \pm 0.13$& $7.08$\\ 
J020540.72-173252.4& $87\pm4$& $0.10$& $6.0\pm0.2$& $9.9\pm0.3$& $7.98 \pm 0.13$& $14 \pm 35$& $7.39 \pm 0.55$& $8.94$\\ 
J020601.09+271211.4& $128\pm15$& $0.14$& $8.3\pm0.5$& $9.0\pm0.5$& $7.73 \pm 0.16$& $34 \pm 26$& $6.78 \pm 0.32$& $8.05$\\ 
J020613.35-060948.8& $83\pm9$& $0.00$& $6.1\pm0.5$& $6.6\pm0.5$& $7.63 \pm 0.21$& $31 \pm 9$& $6.56 \pm 0.08$& $7.79$\\ 
J022225.97-035509.9& $99\pm9$& $0.12$& $5.8\pm0.4$& $8.7\pm0.4$& $7.71 \pm 0.08$& $10 \pm 2$& $6.08 \pm 0.05$& $7.91$\\ 
J024441.11+284600.7& $167\pm31$& $0.05$& $12.5\pm1.5$& $7.6\pm0.5$& $7.71 \pm 0.06$& $11 \pm 5$& $6.53 \pm 0.23$& $8.35$\\ 
J024758.00+035025.0& $145\pm22$& $0.10$& $3.6\pm0.4$& $5.4\pm0.3$& $7.58 \pm 0.20$& $13 \pm 6$& $4.86 \pm 0.12$& $6.50$\\ 
J030024.01+020628.8& $39\pm2$& $0.20$& $1.9\pm0.1$& $8.5\pm0.4$& $7.95 \pm 0.15$& $33 \pm 17$& $6.48 \pm 0.13$& $8.96$\\ 
J031911.82+401729.7& $122\pm17$& $0.16$& $7.6\pm0.7$& $6.7\pm0.4$& $7.53 \pm 0.14$& $15 \pm 5$& $6.18 \pm 0.11$& $7.56$\\ 
J043015.89+084530.5& $144\pm11$& $0.12$& $9.8\pm0.3$& $8.6\pm0.2$& $7.93 \pm 0.05$& $8 \pm 0$& $6.09 \pm 0.05$& $7.68$\\ 
J044412.94-045819.9& $89\pm4$& $0.04$& $6.5\pm0.2$& $10.0\pm0.2$& $8.12 \pm 0.07$& $30 \pm 12$& $7.06 \pm 0.23$& $8.56$\\ 
J051134.12-003818.8& $122\pm9$& $0.08$& $4.8\pm0.2$& $7.8\pm0.3$& $7.69 \pm 0.12$& $36 \pm 16$& $6.60 \pm 0.19$& $7.78$\\ 
J061046.04+644950.1& $110\pm5$& $0.11$& $2.8\pm0.1$& $7.4\pm0.1$& $7.95 \pm 0.08$& $26 \pm 9$& $6.64 \pm 0.11$& $7.93$\\ 
J064101.26+381532.9& $92\pm6$& $0.12$& $4.5\pm0.3$& $6.1\pm0.2$& $7.57 \pm 0.13$& $11 \pm 10$& $6.15 \pm 0.22$& $8.67$\\ 
J075733.93+475030.7& $133\pm8$& $0.08$& $6.2\pm0.3$& $6.0\pm0.2$& $7.50 \pm 0.23$& $24 \pm 9$& $5.93 \pm 0.11$& $7.13$\\ 
J081713.08+263350.3& $63\pm7$& $0.07$& $3.3\pm0.2$& $5.8\pm0.4$& $7.59 \pm 0.36$& $92 \pm 78$& $6.68 \pm 0.38$& $8.39$\\ 
J082225.73+460522.8& $105\pm17$& $0.00$& $9.2\pm0.8$& $8.0\pm0.6$& $7.67 \pm 0.07$& $25 \pm 11$& $5.86 \pm 0.11$& $7.03$\\ 
J083609.47+030600.0& $15\pm1$& $0.00$& $5.1\pm0.2$& $7.3\pm0.2$& $7.54 \pm 0.10$& $7 \pm 4$& $5.23 \pm 0.10$& $7.47$\\ 
J084425.44-024001.1& $92\pm4$& $0.00$& $4.3\pm0.1$& $7.2\pm0.2$& $7.65 \pm 0.11$& $57 \pm 25$& $6.03 \pm 0.10$& $7.23$\\ 
J084530.80+013151.2 $\ast$ & $296\pm63$& $0.12$& $8.8\pm1.1$& $4.5\pm0.2$& $7.30 \pm 0.13$& $2 \pm 1$& $4.73 \pm 0.08$& $6.44$\\ 
J090541.82+253227.9& $48\pm3$& $0.07$& $2.6\pm0.1$& $4.9\pm0.2$& $7.47 \pm 0.13$& $56 \pm 34$& $6.50 \pm 0.18$& $7.88$\\ 
J090700.00-000251.2& $57\pm3$& $0.09$& $4.2\pm0.2$& $8.8\pm0.5$& $7.71 \pm 0.10$& $11 \pm 3$& $6.69 \pm 0.19$& $8.64$\\ 
J091737.69+505715.6& $64\pm5$& $0.00$& $4.6\pm0.4$& $5.8\pm0.3$& $7.60 \pm 0.11$& $64 \pm 37$& $6.22 \pm 0.15$& $7.28$\\ 
J092713.61+031424.6& $87\pm5$& $0.05$& $4.7\pm0.2$& $5.2\pm0.2$& $7.57 \pm 0.10$& $12 \pm 3$& $4.83 \pm 0.06$& $6.98$\\ 
J093751.74+204653.2& $84\pm10$& $0.09$& $3.8\pm0.3$& $5.1\pm0.3$& $7.50 \pm 0.16$& $14 \pm 7$& $5.62 \pm 0.12$& $8.57$\\ 
J095506.70-031617.7& $77\pm4$& $0.10$& $4.9\pm0.2$& $7.7\pm0.3$& $7.76 \pm 0.08$& $184 \pm 130$& $7.50 \pm 0.44$& $8.25$\\ 
J100008.52+273614.9& $110\pm17$& $0.00$& $2.3\pm0.1$& $4.7\pm0.3$& $7.60 \pm 0.24$& $36 \pm 16$& $5.91 \pm 0.11$& $7.11$\\ 
J100438.64+291146.2& $35\pm2$& $0.06$& $3.4\pm0.2$& $5.8\pm0.3$& $7.58 \pm 0.15$& $81 \pm 49$& $6.91 \pm 0.28$& $8.10$\\ 
J100512.15+372201.5& $67\pm4$& $0.00$& $2.2\pm0.1$& $2.7\pm0.1$& $7.25 \pm 0.22$& $60 \pm 32$& $3.65 \pm 0.13$& $4.67$\\ 
J101818.56+213736.6& $74\pm8$& $0.11$& $3.1\pm0.2$& $6.4\pm0.4$& $7.58 \pm 0.08$& $38 \pm 23$& $6.39 \pm 0.16$& $7.71$\\ 
J102500.91+572129.9& $62\pm4$& $0.04$& $3.3\pm0.2$& $5.4\pm0.3$& $7.57 \pm 0.06$& $66 \pm 28$& $6.19 \pm 0.11$& $7.26$\\ 
J103129.68+383210.9& $36\pm2$& $0.11$& $2.0\pm0.1$& $6.4\pm0.3$& $7.71 \pm 0.06$& $115 \pm 57$& $6.82 \pm 0.15$& $7.95$\\ 
J103207.91+562129.6& $44\pm2$& $0.22$& $1.6\pm0.0$& $8.1\pm0.3$& $7.89 \pm 0.06$& $12 \pm 5$& $6.95 \pm 0.21$& $9.76$\\ 
J103400.94-022158.9& $61\pm3$& $0.14$& $2.6\pm0.1$& $6.9\pm0.3$& $7.72 \pm 0.09$& $90 \pm 65$& $7.18 \pm 0.34$& $8.33$\\ 
J103641.07+140715.1& $41\pm2$& $0.10$& $1.9\pm0.1$& $6.0\pm0.2$& $7.59 \pm 0.10$& $22 \pm 11$& $5.76 \pm 0.11$& $7.70$\\ 
J110431.76-004146.5& $111\pm8$& $0.11$& $4.6\pm0.2$& $9.8\pm0.5$& $8.06 \pm 0.12$& $29 \pm 18$& $7.09 \pm 0.30$& $8.36$\\ 
J110521.68+175417.8& $59\pm5$& $0.00$& $4.2\pm0.2$& $7.4\pm0.5$& $7.75 \pm 0.10$& $80 \pm 44$& $6.74 \pm 0.16$& $7.77$\\ 
J110849.73+145358.5& $92\pm6$& $0.04$& $3.1\pm0.1$& $6.4\pm0.2$& $7.59 \pm 0.08$& $27 \pm 8$& $6.28 \pm 0.08$& $7.92$\\ 
J111857.75+670953.3& $62\pm3$& $0.03$& $11.1\pm0.5$& $9.8\pm0.4$& $7.90 \pm 0.05$& $580 \pm 255$& $6.90 \pm 0.16$& $7.57$\\ 
J112712.36+391520.9& $116\pm12$& $0.06$& $3.0\pm0.2$& $6.2\pm0.3$& $7.59 \pm 0.29$& $62 \pm 30$& $6.06 \pm 0.11$& $7.84$\\ 
J113058.67+234836.9& $108\pm7$& $0.03$& $6.4\pm0.3$& $6.4\pm0.3$& $7.58 \pm 0.07$& $19 \pm 3$& $6.11 \pm 0.06$& $7.35$\\ 
J120459.62+525934.6& $64\pm6$& $0.06$& $3.3\pm0.2$& $6.5\pm0.4$& $7.59 \pm 0.11$& $82 \pm 48$& $6.48 \pm 0.20$& $7.45$\\ 
J120652.26+173942.5& $129\pm6$& $0.20$& $6.9\pm0.2$& $8.8\pm0.2$& $7.81 \pm 0.04$& $45 \pm 29$& $6.39 \pm 0.13$& $7.40$\\ 
J122641.86+223034.2& $95\pm6$& $0.07$& $3.9\pm0.2$& $4.5\pm0.1$& $7.46 \pm 0.10$& $12 \pm 3$& $5.93 \pm 0.05$& $7.19$\\ 
J123034.76+370821.3& $58\pm4$& $0.14$& $2.9\pm0.2$& $7.6\pm0.4$& $7.81 \pm 0.15$& $74 \pm 33$& $7.02 \pm 0.12$& $8.04$\\ 
J123920.18+392104.5& $123\pm14$& $0.06$& $7.8\pm0.4$& $6.8\pm0.4$& $7.50 \pm 0.06$& $11 \pm 5$& $6.56 \pm 0.21$& $8.15$\\ 
J125134.65+392642.6& $94\pm7$& $0.04$& $9.0\pm0.8$& $5.8\pm0.3$& $7.51 \pm 0.19$& $42 \pm 28$& $5.80 \pm 0.12$& $6.94$\\ 
J125727.52+270605.3& $77\pm8$& $0.09$& $3.8\pm0.2$& $6.7\pm0.4$& $7.58 \pm 0.11$& $40 \pm 34$& $6.21 \pm 0.18$& $7.40$\\ 
\hline
\end{tabular}
\end{table*}

\section{Results}
\label{sec:results}

We developed a method for identifying XMPs hosting prominent young stellar populations with ground-based \uband{}\gband{}\rband{}\iband{}\zband{} photometric data.
Applying this technique to the SDSS photometric database yielded over 150 candidates not observed in the SDSS spectroscopic survey.
As part of an ongoing spectroscopic follow-up survey, we obtained confirmation spectra with MMT Red and Blue Channel, yielding direct-$T_e$ metallicities for 53 new systems.
Here we present the results of these first spectroscopic observations in three sections, focusing on the broadband photometry (Section~\ref{sec:res_phot}), the gas-phase metallicities (\ref{sec:res_met}), and the ionization state of the gas (\ref{sec:res_gasion}).

\subsection{Broadband photometric properties}
\label{sec:res_phot}

The galaxies selected for spectroscopic follow-up in this paper have median apparent magnitudes $\uband{}=19.3$ and $\iband{}=19.0$, significantly below the completeness limit of the SDSS spectroscopic sample \citep[$\rband{}\lesssim 17.8$:][and evident in Figure~\ref{fig:magdist_select}]{Strauss2002}.
They are also quite compact, with median effective radius 1.4$''$ in the \iband{}-band (as reported by the SDSS photometric pipeline from exponential profile fits accounting for the point spread function; Table~\ref{tab:targs}).
At the spectroscopic distances of these galaxies, the measured effective radii correspond to physical sizes of $0.4^{+0.6}_{-0.2}$ kpc (16-50-84 percentile range).
This is comparable to the typical measured sizes of $z>6$ star-forming galaxies \citep[e.g.][]{Shibuya2015,Curtis-Lake2016}.

\begin{figure}
    \includegraphics[width=0.5\textwidth]{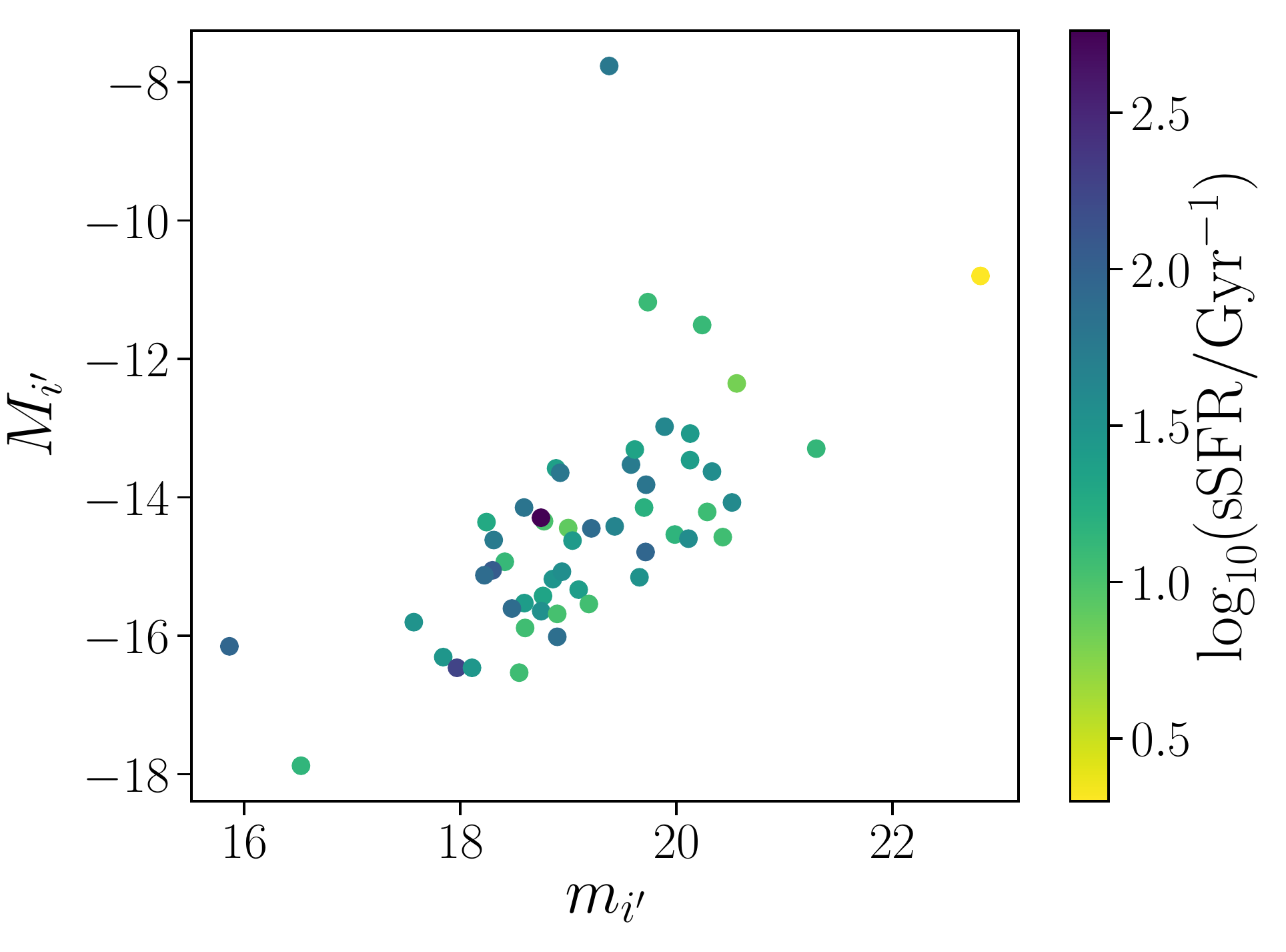}
    \caption{
        Absolute (corrected for redshift-derived distances) versus apparent $i$-band magnitudes for our sample (from the SDSS \texttt{modelMag} measurements), color-coded by the specific star formation rates inferred from our SED fitting.
        The formal uncertainties in the photometry are typically smaller than the point sizes used here, and thus are not plotted.
        The galaxies presented in this paper are almost all fainter than the completeness limits of the SDSS spectroscopic survey ($\rband{}<18$), extending down to $\iband{} = 23$.
        Due to our selection based on band contamination in both the \gband{} and \rband{} bands, our sample is effectively limited to finding objects within 130 Mpc, and thus apparently fainter objects selected for follow-up tend to be intrinsically fainter on average as well.
        The majority of our targets have faint absolute magnitudes $M_{\iband{}}>-15$, within the range in which we expect to encounter extremely metal-poor gas \citep[e.g.][]{Berg2012}.
    }
    \label{fig:appmagabsmag}
\end{figure}

Our aim is to identify galaxies at very low metallicities, which likely have relatively low stellar masses \citep[$M/M_\odot \lesssim 10^{7.5}$,][]{Berg2012}.
In Figure~\ref{fig:appmagabsmag} we plot the apparent magnitude of the targeted galaxies against the absolute magnitude derived with our spectroscopic redshifts.
The galaxies we targeted are uniformly intrinsically faint, with median absolute magnitude of $-14.6$ in the \iband{} band.
Similarly, all have stellar masses derived from constant star formation model fits below $M/M_\odot = 10^{7.5}$ and extend down to $M/M_\odot = 10^{3.7}$, with median $M/M_\odot = 10^{6.3}$ (Table~\ref{tab:linemet}).
Our total stellar mass estimates computed with the color-dependent mass-to-light ratios provided by \citet{Bell2003} yield a median mass of $M/M_\odot = 10^{7.8}$ (Section~\ref{sec:phot}), which in the context of the mass--metallicity relation found by \citet{Berg2012} with masses computed in a similar manner suggests we are in the regime of very low-metallicity systems.
The faintest systems in our sample also tend to have fainter absolute magnitudes and consequently lower stellar masses (Figure~\ref{fig:appmagabsmag}).
The galaxies fainter than $\iband{}\simeq 20$ have a median absolute magnitude of $M_{\iband}=-13.5$, 1.5 magnitudes fainter than galaxies brighter than this cutoff.
This is a result of our selection technique identifying systems only at low redshifts $z<0.03$ where both the $g$ and $r$ band are contaminated (Section~\ref{sec:photsamp} and Table~\ref{tab:targs}).
Since our search is thus effectively limited to a 130 Mpc radius volume, restricting a subsample to apparent magnitudes fainter than some limit is equivalent to placing an upper bound on intrinsic luminosity and thus stellar mass.

The broadband photometry also constrains the stellar populations present in these systems.
Figure~\ref{fig:appmagabsmag} displays the specific star formation rates derived from the constant star formation history SED fits in the marker colors.
The median sSFR in our sample is 36 \unit{Gyr^{-1}}, which indicates that the optical SEDs of these galaxies are predominantly shaped by stellar populations formed within the last 30 Myr (Table~\ref{tab:linemet}).
This is similar to values measured in a comparable way in the reionization era \citep[e.g.][and references therein]{Stark2016}, and substantially higher than in most known XMPs (see Section~\ref{sec:res_gasion}).
Consider the system J0822+4605, a galaxy with stellar mass in its recent burst of $M/M_\odot = 10^{5.9\pm 0.1}$ and sSFR $40\pm 18$ \unit{Gyr^{-1}} typical of our sample.
Despite its relatively low stellar mass in stars produced recently, this system has had sufficient recent star formation to produce a significant population of hot stars.
Under nominal assumptions ($Z/Z_\odot=0.1$ and the fiducial BPASS IMF, which is approximately \citealt{Kroupa1993}), the BPASS v2.0 stellar synthesis models \citep{Eldridge2017} indicate that this star formation history corresponds to $\sim 1000$ active O stars.
This galaxy has a measured effective radius of 1.3\arcsec{} at 52 Mpc distant (Figure~\ref{fig:cutouts}), suggesting that this substantial young stellar population occupies a region $\sim 33$ pc in radius.
The bulk properties of the dense burst in J0822+4605 are comparable to individual super star clusters observed in the canonical XMP SBS 0335-052, which is similarly undergoing a particularly intense phase of star formation \citep{Reines2008}.

\subsection{Metallicities}
\label{sec:res_met}

Our photometric selection was designed to locate nearby star-forming dwarf galaxies at extremely low metallicity ($Z/Z_\odot < 0.1$).
The photometric data and spectroscopic redshifts obtained for the set of candidates presented in this paper confirm that we have identified nearby $<130$ Mpc compact galaxies with optical SEDs dominated by low-mass $M/M_\odot = 10^{3.7}$--$10^{7.5}$ young stellar populations (Section~\ref{sec:res_phot}).
Now we examine the gas-phase metallicities measured for this sample.

\begin{figure}
    \includegraphics[width=0.5\textwidth]{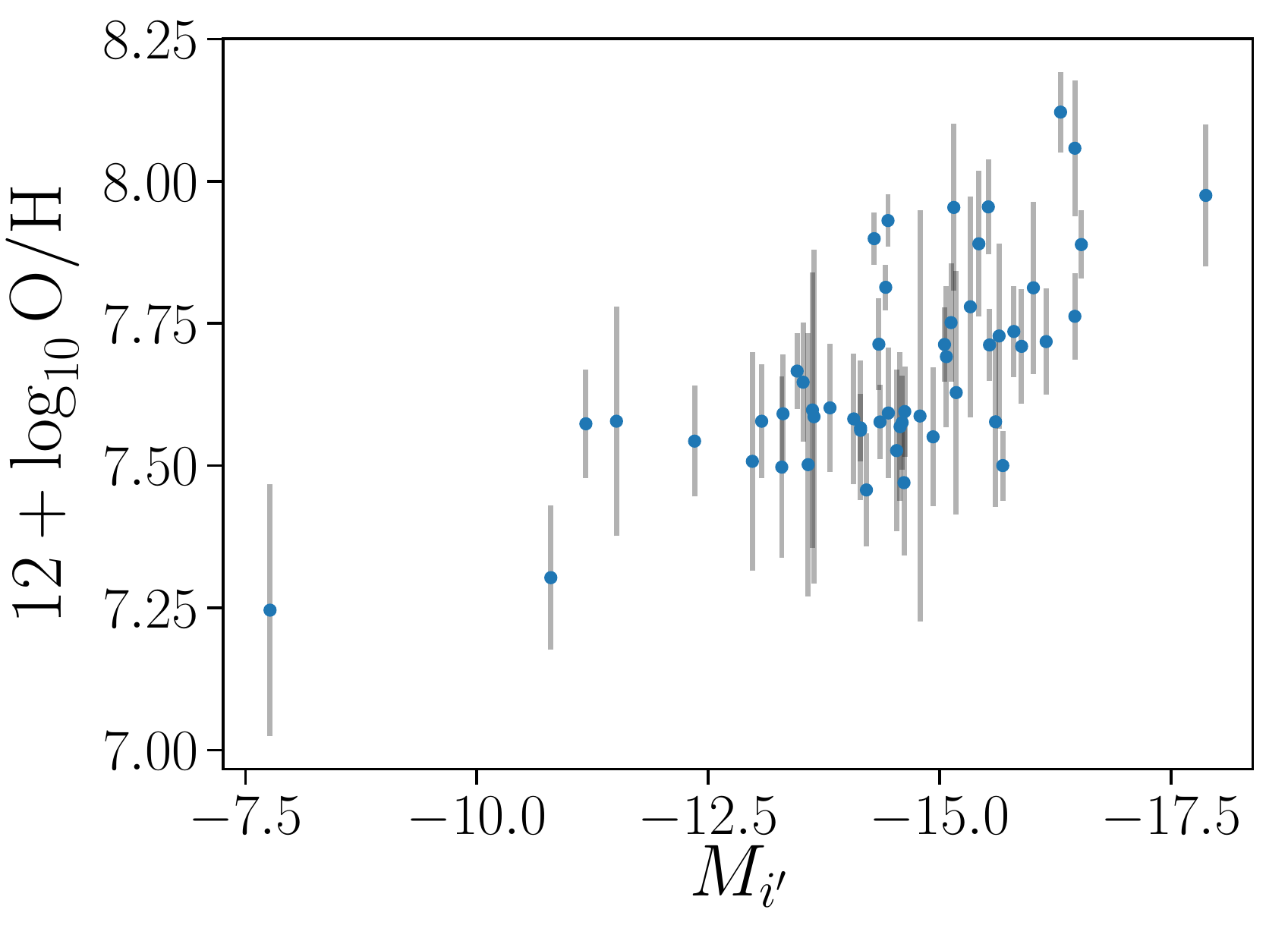}
    \caption{
        Metallicity versus absolute magnitude for the newly-confirmed galaxies presented in this paper.
        Of the 53 photometric targets we obtained spectra for, 32 revealed extremely metal-poor gas $12+\log\mathrm{O/H}<7.7$.
        The absolute magnitudes of our targets are uniformly faint $M_{\iband{}}>-18$, but we still observe a significant trend between these parameters.
        The intrinsically faint distribution is heavily weighted towards low metallicities, with 28/32 fainter than $M_{\iband{}}>-15$ classifying as XMPs.
    }
    \label{fig:absmagmet}
\end{figure}

The metallicity distribution of our sample is displayed in Figure~\ref{fig:absmagmet}, where we plot the direct-$T_e$ oxygen abundances against absolute magnitude in the continuum \iband{}-band for our sample (see also Table~\ref{tab:linemet}).
Of the 53 systems for which we measured [\oiii{}] $\lambda 4363$ and derived metallicities, over half (32) fall below $12+\log\mathrm{O/H}=7.7$ and are thus classified as XMPs, with all but 2 of the others falling below $12+\log\mathrm{O/H}= 8.0$ ($Z/Z_\odot\lesssim 0.2$).
This is a higher fraction of XMPs than found in other photometrically-selected nearby galaxy samples \citep[e.g.][]{Brown2008,Amorin2010,Yang2017}.
For example, the Green Peas are selected on strong [\oiii{}]/H$\alpha$ emission (see Section~\ref{sec:photsamp}), and as a result span the range $7.7 \lesssim 12+\log\mathrm{O/H} \lesssim 8.4$ with mean $12+\log\mathrm{O/H} = 8.05\pm 0.14$ \citep{Amorin2010}; essentially none fall into the range of XMPs.
This confirms that two-band excess selections applied at faint magnitudes are extraordinarily efficient at identifying nearby XMPs.

Galaxies below $Z/Z_\odot = 0.05$ remain particularly elusive \citep[e.g.][]{Guseva2017,Hsyu2017,Izotov2018}, and locating populations of massive stars in this regime is thus extremely valuable for testing stellar models.
Among the 32 confirmed XMPs in our sample, we have identified two new systems in this regime:
J0845+0131 at $12+\log\mathrm{O/H}=7.30\pm0.13$ and J1005+3722 with $7.25\pm0.22$ are compact systems in the realm of the lowest-metallicity galaxies discovered to-date.
Though both are characterized by prominent massive star populations, J0845+0131 stands out, with a dominant young stellar population likely formed within the last several Myr and as a result the highest equivalent width H$\beta$ emission in our sample ($296\pm 63$ \AA{}).
This system is one of only three known at $12+\log\mathrm{O/H}\leq 7.30$ with H$\beta$ equivalent width $\geq 300$ \AA{}, the other two being one of the star-forming regions in SBS0335-052E and J0811+4730 \citep{Izotov2009,Guseva2017,Izotov2018}.
Deeper spectroscopy of these systems will allow detailed investigation of massive star populations and gas conditions in this relatively unexplored regime of extremely young and metal-poor systems.

The lowest-metallicity galaxies in our sample present the faintest absolute magnitudes and lowest stellar masses as well, with J0845+0131 at $M_{\iband{}}=-10.8$ and $10^{4.73\pm0.08} M_\odot$; and J1005+3722 at $M_{\iband{}}=-7.8$ and $10^{3.65\pm 0.13} M_\odot$ in recently-formed stars.
Adopting large aperture photometry and assuming a higher mass-to-light ratio (Section~\ref{sec:phot}) still yields very low total stellar masses of $10^{6.4}$ and $10^{4.7}$ $M_\odot$, respectively.
Due to its very low redshift $z=0.0013$, the mass of J1005+3722 is subject to large systematic uncertainties from our adopted local velocity flow model; but even a factor of two underestimate leaves its total stellar mass $\leq10^5 M_\odot$ and distance $<6$ Mpc.
This places it as potentially one of the nearest known population of massive stars in this metallicity regime $Z/Z_\odot<0.05$, at a similar distance as the gas-rich XMPs Leo P \citep[$\lesssim 2.0$ Mpc;][]{Skillman2013} and the Leoncino Dwarf \citep[7--20 Mpc;][]{Hirschauer2016} discovered through \hi{} surveys.
Systems this nearby can be partially-resolved into individual stars or small clusters, enabling even more detailed investigation of their stellar properties with deep spectroscopy and imaging than for comparable galaxies beyond $>10$ Mpc.

As we are interested in constraining the properties of the lowest-metallicity stars, a technique to distinguish these galaxies at $<0.05$ $Z_\odot$ from those at $0.1$ $Z\odot$ photometrically would be extremely valuable.
Our initial selection employed a blue $\gband{}-\rband{}$ cut to select galaxies with similar contamination in both \gband{} from [\oiii{}]+H$\beta$ and \rband{} from H$\alpha$ (Section~\ref{sec:photsamp}), and correspondingly low [\oiii{}]/H$\alpha$ ratios.
Since this color cut allowed us to efficiently identify systems with $12+\log\mathrm{O/H}<7.7$, we might expect this color to be a useful way to identify systems at the very lowest metallicities with $12+\log\mathrm{O/H}<7.35$.
However, the SDSS $\gband{}-\rband{}$ color appears to be a poor predictor of metallicity in the blue ($\gband{}-\rband{}>-0.15$) and faint ($\iband{}>19$) range of our sample.
We see little correlation between gas-phase metallicity and $\gband{}-\rband{}$ color among galaxies that satisfy our initial selection, and the two lowest-metallicity systems at $12+\log\mathrm{O/H}<7.35$ fall near the median of our sample in $\gband{}-\rband{}$ (at -0.04 and 0.03; Tables\ref{tab:targs} and \ref{tab:linemet}).
Photometric uncertainties and redshift-bandpass interactions likely increase the scatter in the relationship between $\gband{}-\rband{}$ and [\oiii{}]/H$\alpha$, making SDSS colors alone a poor indicator of metallicity in this domain.

However, we do observe a significant trend between absolute magnitude and metallicity in this sample (Figure~\ref{fig:absmagmet}).
The galaxies in our sample with absolute magnitudes $M_{\iband{}}>-15$ are significantly more metal-poor than those brighter than this cutoff.
Among these faintest systems, 28 of 32 are XMPs with $12+\log\mathrm{O/H}<7.7$, whereas only 4 of the 21 intrinsically brighter systems fall into this category.
This trend can be understood as a manifestation of the well-known mass--metallicity relationship \citep[e.g.][]{Berg2012}, and is clear also in examining our stellar mass estimates (Table~\ref{tab:linemet}).
While absolute magnitude and mass determination requires spectroscopic redshift measurement, our sample effectively identifies only galaxies within 130 Mpc.
We might then expect apparent magnitudes to correlate with metallicity as well.

\begin{figure}
    \includegraphics[width=0.5\textwidth]{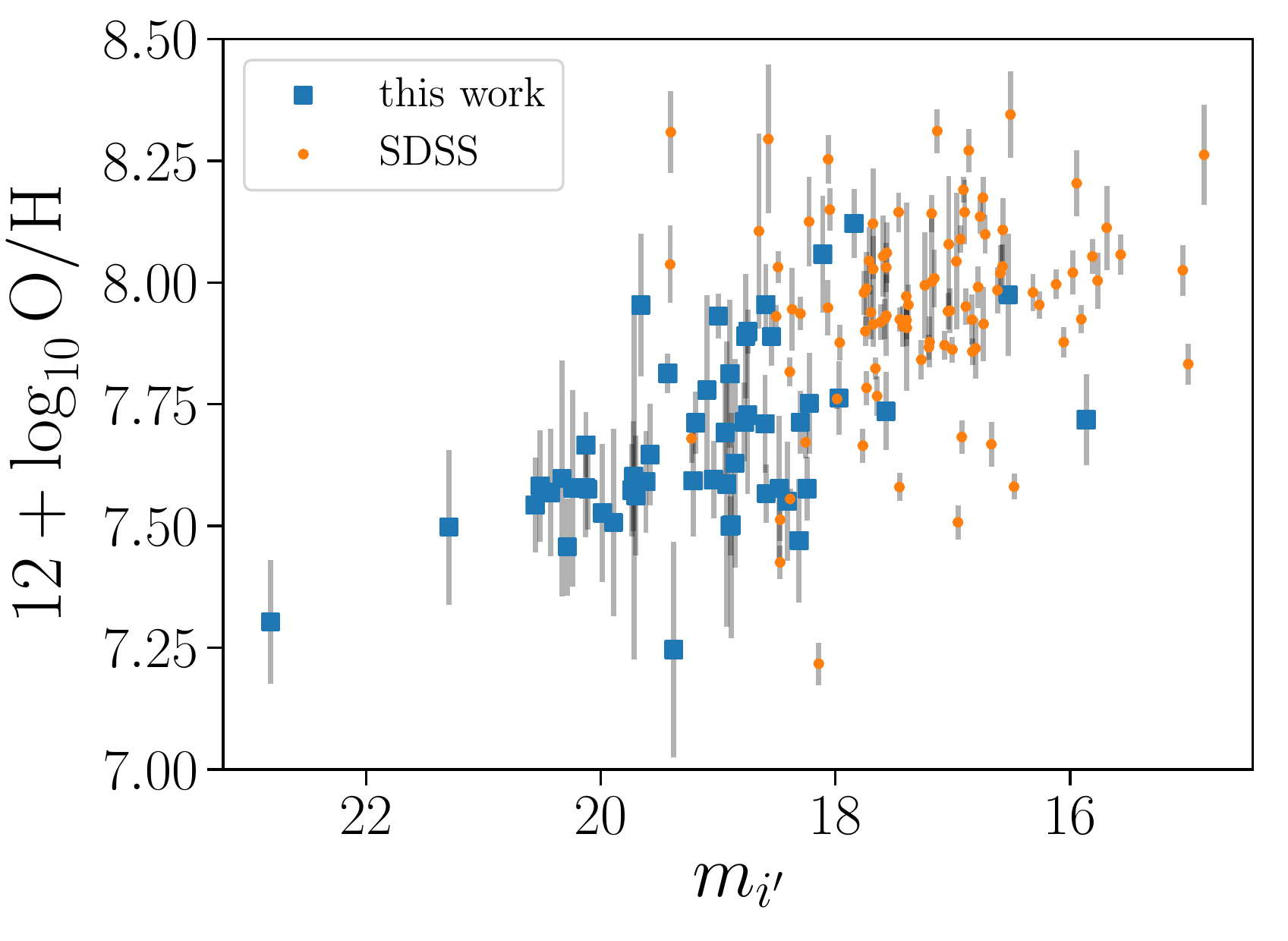}
    \caption{
        The apparent \iband{}-band continuum magnitude versus metallicity distribution for objects selected by our photometric technique.
        We include 96 objects with existing SDSS spectra for which we measured metallicities using the technique described in Section~\ref{sec:specobs}, but with [\oii{}] $\lambda \lambda 7320,7330$ in cases where the [\oii{}] $\lambda\lambda 3727,3729$ doublet was missed by the SDSS spectrum (Section~\ref{sec:nebmet}).
        The galaxies with spectra presented for the first time in this paper are represented by blue squares.
        These objects largely fall below the apparent magnitude completeness limit of the SDSS spectroscopic sample, and have significantly lower typical metallicities.
    }
    \label{fig:appmagmet}
\end{figure}

In Figure~\ref{fig:appmagmet} we plot the metallicity of objects selected by our photometric technique against their apparent \iband{}-band (continuum) magnitude, including brighter candidates which were allocated SDSS fibers.
As described in Section~\ref{sec:nebmet}, metallicities were measured in the SDSS spectra and our MMT spectra in a self-consistent manner.
This plot reveals a significant correlation between apparent magnitude and metallicity.
Galaxies fainter than $m_{\iband{}}=19$ are systematically lower-metallicity than those brighter, with a median metallicity of $12+\log\mathrm{O/H}=7.59$ that is well into the XMP regime at $m_{\iband{}}>19$ compared to the median of $12+\log\mathrm{O/H} = 7.96$ in the brighter systems.
The two lowest-metallicity galaxies in our sample at $12+\log\mathrm{O/H}<7.35$ are also the most intrinsically-faint, at $M_{\iband}=-10.8$ and $-7.8$; and only one (J1005+3722) was sufficiently nearby to reach an apparent magnitude of $m_{\iband{}}=19.4$ and be resolved by SDSS.
This suggests that focusing this technique on fainter objects using deeper surveys like LSST and HSC will yield large numbers of galaxies at or below the metallicity floor of current XMP samples, substantially expanding the sample of systems to which deep spectroscopy can be applied to understand the lowest-metallicity massive stars.

\subsection{Gas properties and ionizing spectra}
\label{sec:res_gasion}

We have identified 32 new XMPs selected using a photometric band excess technique.
These systems are systematically fainter than XMPs discovered by the magnitude-limited SDSS spectroscopic sample.
By virtue of their selection on high equivalent width nebular emission and volume-limited distribution, the XMPs presented here uniformly host prominent massive star populations (median effective stellar population age of $28$ Myr) in a compact region $<1$ kpc in radius (Section~\ref{sec:res_phot}).
Future deep spectroscopy of these objects will enable stellar features such as Wolf-Rayet wind emission to be constrained.
However, our discovery spectra already allow us to investigate the bulk gas conditions and to address the still-mysterious origin of high-ionization nebular \heii{} emission in these systems.

\begin{figure}
    \includegraphics[width=0.5\textwidth]{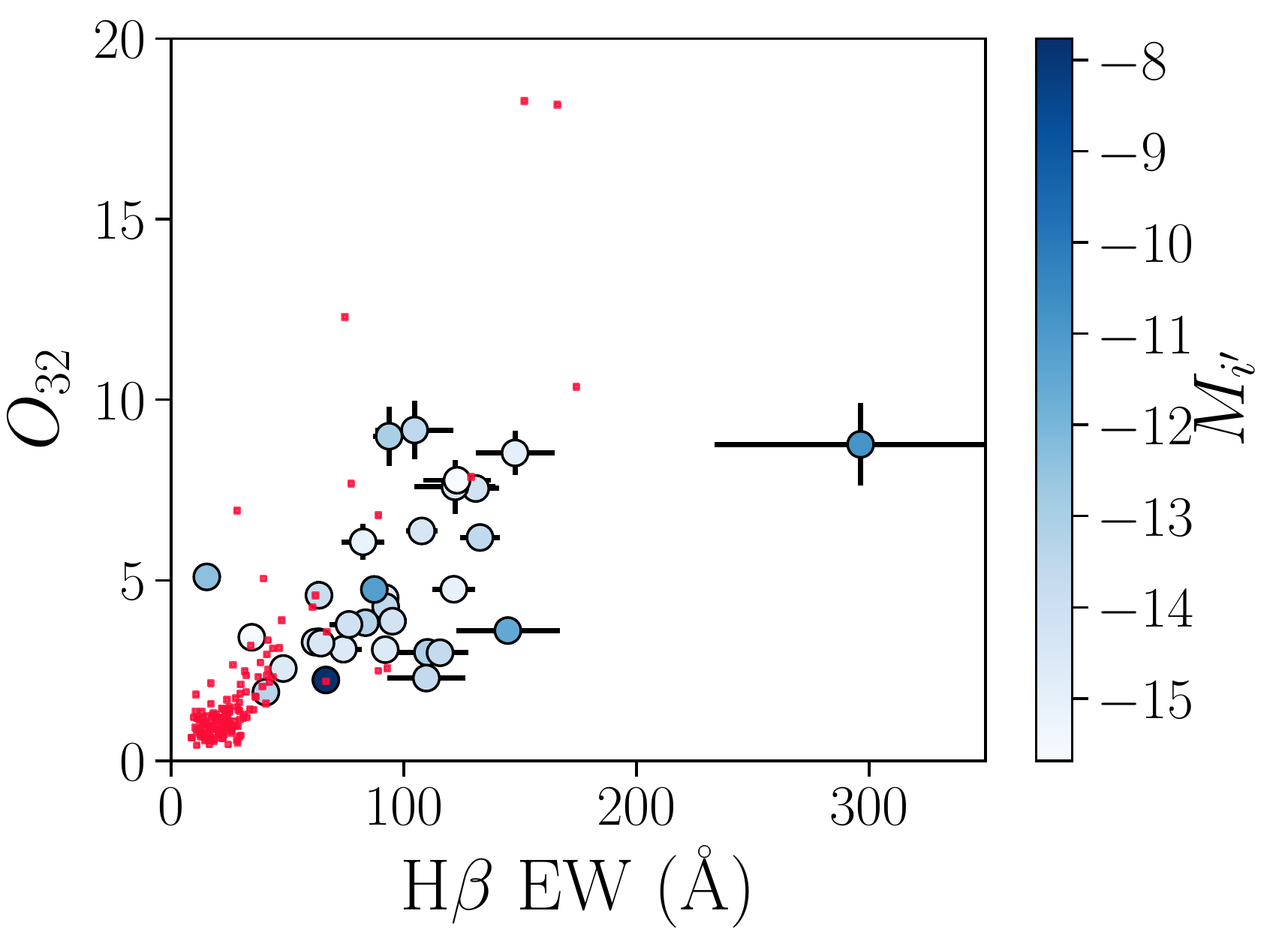}
    \caption{
        H$\beta$ equivalent widths (a proxy for sSFR) versus dust-corrected \ott{} values (a proxy for ionization parameter) for our galaxies (blue circles), color-coded by absolute \iband{}-band magnitude.
        We plot as black points the XMPs uncovered uniformly from the SDSS spectroscopic sample by \citet{SanchezAlmeida2016}.
        Uncertainties ($1\sigma$) are plotted for our galaxies, and are hidden by the markers in several cases.
        The vast majority of SDSS XMPs are clustered at H$\beta$ EWs $<50$ \AA{} and \ott $<3$.
        Relatively few previously-identified XMPs extend into the high-sSFR region of this diagram at H$\beta$ EWs $>50$ where our objects reside.
        The gas in the XMPs presented here is also more highly-ionized than in most XMPs from the SDSS spectroscopic sample, consistent with substantially different stellar populations and resulting ionizing radiation fields.
    }
    \label{fig:o32_hbeta}
\end{figure}

We first explore how the recent star formation histories and ionized gas conditions in our galaxies compare to those in brighter XMPs discovered previously.
In Figure~\ref{fig:o32_hbeta}, we plot H$\beta$ equivalent widths and \ott{} ratios for the XMPs presented in this paper.
For comparison, we also plot XMPs identified by \citet{SanchezAlmeida2016} from the SDSS spectroscopic survey.
Line measurements and dust corrections for the \citet{SanchezAlmeida2016} objects are conducted in the same manner as for our sample (Section~\ref{sec:nebmet}); as we are plotting \ott{}, we only include those systems with [\oii{}] $\lambda \lambda 3727,3729$ constraints from SDSS.
Reionization-era galaxies are expected to present both high \ott{} \citep[e.g.][]{Nakajima2016,Tang2018_arxiv} and high equivalent-width H$\beta$.
Since the equivalent width of H$\beta$ is anticorrelated with the age of a simple or constantly star-forming stellar population, this quantity is a useful empirical proxy for the light-weighted stellar age of a galaxy.
This plot reveals that the bulk of XMPs identified from SDSS spectra present relatively low H$\beta$ equivalent widths.
The median H$\beta$ equivalent width of our sample of XMPs is 92 \AA{}, three times larger than the median for the SDSS XMPs of 28 \AA{}.
This difference suggests that the XMPs we have identified here are dominated by much younger stars compared to typical previously-known XMPs, presenting sSFRs and effective ages closer to those expected in early galaxies.
As a result of the relative preponderance of young massive stars in these systems, we might expect the gas in these objects to show evidence for a more intense overall ionizing radiation field.

The \ott{} ratio is commonly used as a proxy for the ionization parameter, or the density of ionizing radiation relative to the gas density.
Figure~\ref{fig:o32_hbeta} reveals that indeed, the XMPs presented here host more highly-ionized gas than most XMPs from the SDSS spectroscopic sample.
Our XMPs have \ott{} ratios ranging as high as 10, with median $\ott{}=4.1$ (and an even higher median $\ott{}=6.4$ for the subset with H$\beta$ equivalent widths above 100 \AA{}).
This is substantially higher than the majority of SDSS XMPs shown, which have median $\ott{}=1.2$.
This indicates that the gas in the galaxies we have uncovered is more highly-ionized than in most previously-known XMPs, consistent with these systems hosting harder ionizing radiation fields.

The offset towards higher H$\beta$ equivalent widths and \ott{} ratios in our XMPs suggests that these systems host significantly more prominent young massive stellar populations than typical previously-known XMPs.
Figure~\ref{fig:o32_hbeta} indicates that the average XMP from the SDSS spectroscopic sample has undergone a more extended star formation history, resulting in an optical spectrum with a substantial contribution from stars formed $>10$ Myr ago and less highly-ionized gas.
The EUV and X-ray radiation powered by a typical nearby XMP with a H$\beta$ equivalent width of 30 \AA{} (sSFR$\simeq 1$ \unit{Gyr^{-1}} under constant star formation) is potentially very different from that of a reionization-era galaxy which has formed essentially all of its stars in the last 100 Myr (sSFR $\simeq$ 5--10 \unit{Gyr^{-1}}).
Future studies targeting these high-sSFR XMPs will provide a new window onto the massive stars and X-ray populations that likely dominate the early universe.

\begin{figure}
    \includegraphics[width=0.5\textwidth]{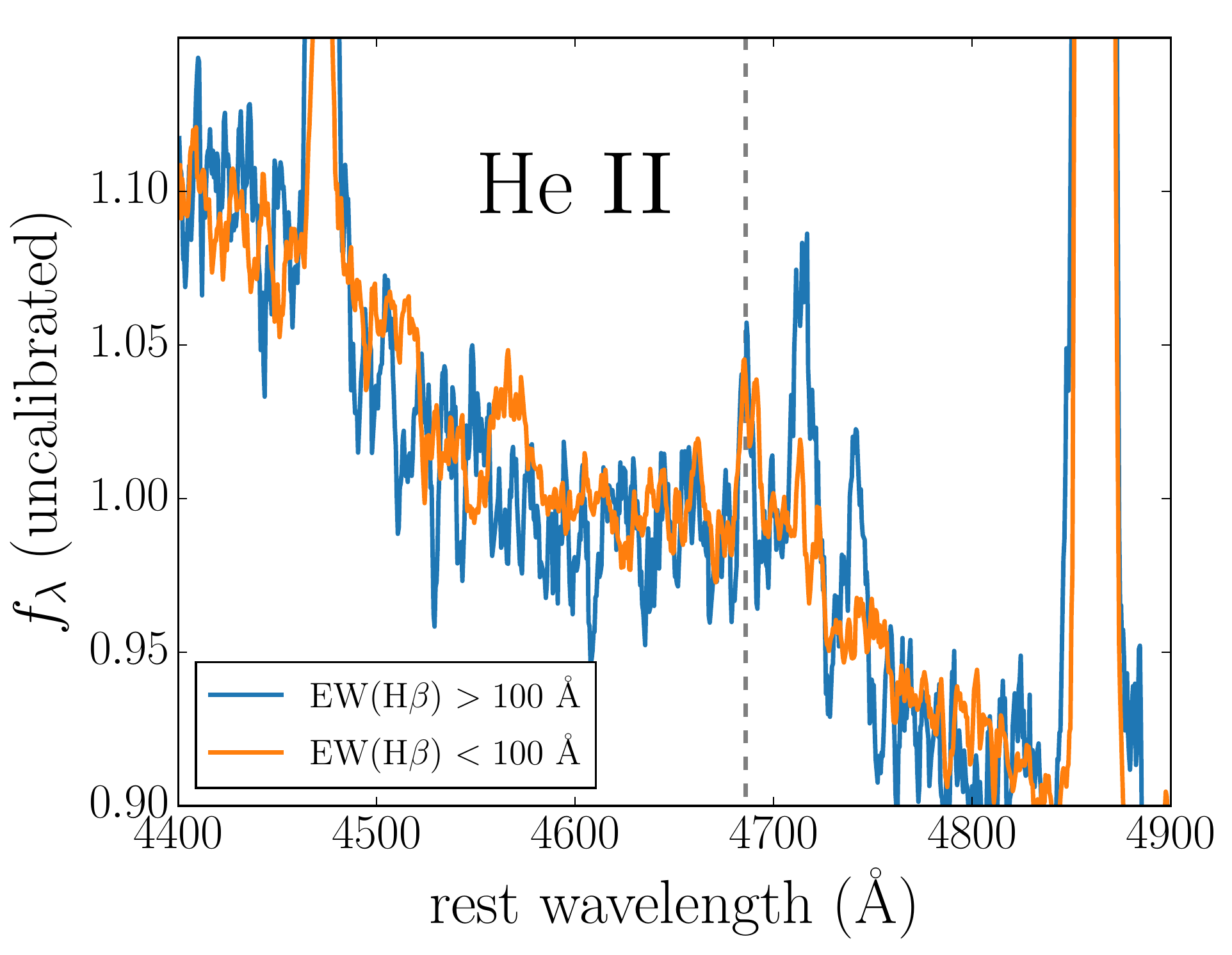}
    \caption{
        The median-stacked continuum normalized spectra of the galaxies in our sample centered on \heii{} and separated into two bins above and below an H$\beta$ equivalent width of 100 \AA{}.
        These stacks both reveal prominent narrow \heii{} $\lambda 4686$ \AA{} emission.
        Surprisingly, the equivalent width of \heii{} is visibly unchanged between the two stacks though the equivalent widths of the strong nebular lines are significantly different.
        This suggests that some of the hard ionizing radiation $>54.4$ eV is decoupled from the lifetimes of the most massive stars ($<10$ Myr), and may require invoking longer-lived stellar sources such as binary evolution products.
        Deeper data and \heii{} detections in a range of individual XMPs will be required to draw firmer conclusions.
    }
    \label{fig:heiistack}
\end{figure}

As a first step towards constraining their EUV ionizing spectra, we examine high-ionization nebular \heii{} emission in our galaxies.
The \heii{} $\lambda 4686$ nebular emission line is a recombination line powered by extremely energetic photons $>54.4$ eV.
This line is commonly observed in low-metallicity star-forming dwarf galaxies \citep[e.g.][]{Garnett1991,Brinchmann2008,Shirazi2012,Senchyna2017}, but the origin of the necessary $>54.4$ eV photons is still a matter of debate.
The commonly-cited candidate sources of narrow \heii{} are very massive stars and the fast radiative shocks their supernovae drive, lower-mass products of binary stellar evolution, and X-ray binary systems.
By median-stacking the MMT spectra presented in this paper, we can constrain the presence of this line in these new galaxies.
Stacking the continuum-normalized MMT spectra reveals prominent narrow \heii{} (Figure~\ref{fig:heiistack}).
We fit the line complex around \heii{} simultaneously with H$\beta$ after subtracting a fit to the continuum using \texttt{pyspeckit} \citep{Ginsburg2011}, simultaneously fitting 6 Gaussians with a common width to \hei{} $\lambda 4471$, [\feiii{}] $\lambda 4658$, \heii{} $\lambda 4686$, [\ariv{}]+\hei{} $\lambda 4712$, [\ariv{}] $\lambda 4740$, and H$\beta$, and measure uncertainties in these line measurements by repeatedly perturbing the stacked spectrum by the $1\sigma$ residual noise and refitting.
The full stack reveals a flux ratio \heii{} $\lambda 4686$ / H$\beta$ $=0.0133 \pm 0.0004$.
Assuming Case B recombination, this ratio is tied to the spectral hardness ratio $\mathrm{Q}(>54.4 \mathrm{ev})/\mathrm{Q}(>13.6 \mathrm{eV})$, and the measured ratio corresponds to $\mathrm{Q}(>54.4 \mathrm{ev})/\mathrm{Q}(>13.6 \mathrm{eV}) \simeq 0.006$; this is extremely difficult for stellar population models to reproduce \citep[e.g.][and references therein]{Shirazi2012}, but is consistent with measurements in other galaxies below $Z/Z_\odot < 0.2$ \citep[e.g.][]{Senchyna2017}.

We can also investigate how this spectral hardness changes with other galaxy parameters such as metallicity by stacking separate bins of our sample.
Stellar population models predict harder ionizing spectra beyond the $\mathrm{He^+}$-ionizing edge with decreasing stellar metallicity, as stars evolve to hotter temperatures and stellar winds diminish in density.
This is observed in local star-forming regions, with typical \heii{}/H$\beta$ ratios increasing by a factor of 5--10 below $Z/Z_\odot \simeq 0.2$ \citep[e.g.][and references therein]{Senchyna2017}.
We first examine two stacks of our sample at gas-phase metallicities above and below $12+\log\mathrm{O/H}=7.6$, measuring \heii{}/H$\beta$ in each.
We find a slightly larger value of $0.0135\pm0.0006$ above $12+\log\mathrm{O/H} =7.65$ than in systems below it ($0.0118\pm 0.0005$), but both are near the value measured in the full stack and are substantially elevated relative to high-sSFR systems above $12+\log\mathrm{O/H}=8.0$.
The small difference observed with decreasing metallicity may reflect the relatively small dynamic range in metallicity spanned by our stacks; the median metallicities of the galaxies in each are $12+\log\mathrm{O/H} = 7.78$ and $7.57$.

Canonical stellar models and the fast radiative shock explanation both predict that $>54.4$ eV photons will be dominated by the most massive stars and their supernovae, and thus that \heii{} should correlate with the presence of very massive stars with lifetimes $<10$ Myr \citep[e.g.][]{Schaerer1998}.
Adopting this picture, we expect to find that \heii{} is strongest in systems with large H$\beta$ equivalent widths indicative of more dominant recent star formation.
We thus consider two median stacks split in H$\beta$ equivalent width above 100 \AA{} (21 objects, median 131 \AA{}) and below (32 objects, median 61 \AA{}) (see Figure~\ref{fig:heiistack}).
This plot shows that the metal lines nearby \heii{} appear to qualitatively scale with H$\beta$, indicating that the ionizing radiation powering them is coupled to the very massive $<10$ Myr stars dominating the \hi{}-ionizing flux.
But surprisingly, we find \heii{} $\lambda 4686$ at nearly identical equivalent width in both stacks ($1.07\pm 0.06$ \AA{} among those with H$\beta$ $>100$ \AA{}, and $0.96 \pm 0.05$ \AA{} in the $<100$ \AA{} stack).
We measure a \heii{}/H$\beta$ ratio of $0.0171\pm 0.0008$ in the H$\beta$ $<100$ \AA{} stack, nearly double that measured for the highest sSFR systems with H$\beta > 100$ \AA{} ($0.0093\pm 0.0005$).
This suggests that a source with characteristic timescales longer than 10 Myr or effectively decoupled from the lifetimes of the most massive stars contributes substantially to the $\mathrm{He^+}$-ionizing photon budget in these systems.
Binary stellar evolution pathways can produce stripped stars from lower-mass progenitors with substantial flux at 54.4 eV on timescales of 20 Myr or greater \citep{Gotberg2017}.
In addition, X-ray binaries with A-type or later donor stars can live for 10--100 Myr and power relatively soft spectra with potentially a more significant impact at 54.4 eV than their high-mass counterparts \citep[e.g.][]{Fabbiano2006}.
Other evolved products of longer-lived stars such as post-AGB stars may also play some role in powering this line \citep[e.g.][]{Binette1994}.
Further constraints on the nature of this ionizing radiation will require deeper spectra of individual XMPs and more detailed model comparison.
However, this simple experiment highlights the importance of examining how integrated galaxy measurements change with stellar population properties when comparing nearby systems to stellar models or to galaxies in the early universe.

\section{Summary and Outlook}
\label{sec:conclude}

We have described a broadband color selection designed to identify low-redshift $z<0.03$ XMPs with prominent recent star formation.
We then presented the first results of a spectroscopic campaign targeting candidates selected using this technique from SDSS imaging, extending 3 magnitudes below the completeness limits of the SDSS spectroscopic sample.
We detected the [\oiii{}] $\lambda 4363$ auroral line in 53 systems, allowing measurement of direct-$T_e$ gas-phase oxygen abundances.
All but one of these systems was found to have $12+\log\mathrm{O/H}<8.0$ ($Z/Z_\odot<0.2$), and more than half are XMPs (with $12+\log\mathrm{O/H}<7.7$).
We discovered 32 galaxies in this metallicity range, including two of the most metal-poor star-forming galaxies known, with $Z/Z_\odot < 0.05$ and very low-mass young stellar populations ($M/M_\odot < 10^5$).

Observations of local XMPs can potentially shed light on the physics of $Z/Z_\odot<0.1$ massive stars and the nature of reionization-era galaxies \citep[e.g.][]{Crowther2006,Brorby2016,James2017}.
However, they are not a uniform population: known XMPs display a variety of star formation histories (Section~\ref{sec:res_gasion}).
The majority of the relatively bright XMPs discovered by SDSS have significantly less dominant massive star populations than inferred for systems at $z>6$, manifesting in low typical H$\beta$ equivalent widths and ionization parameters (Figure~\ref{fig:o32_hbeta}).
The stellar populations and radiation fields in such systems are likely substantially different from galaxies in the reionization era, and as a result typical XMPs cannot be used as analogs of high-redshift galaxies.
The XMPs uncovered with our photometric technique have larger sSFRs and more highly-ionized gas than the bulk of XMPs discovered previously by SDSS, and thus serve as a step towards a more complete laboratory for testing models of extremely metal-poor massive stars.

We demonstrate the importance of considering the varied stellar populations occupying nearby XMPs by examining the mysterious high-ionization \heii{} $\lambda 4686$ line.
We found that unlike the other nebular lines, the equivalent width of \heii{} does not correlate significantly with H$\beta$ equivalent width in our sample, suggesting that some of the extremely high energy $>54.4$ eV flux in these galaxies is produced by sources with timescales $>10$ Myr such as stripped stars produced by close binary evolution or low-mass X-ray binaries.
Examining XMPs with the youngest effective ages (high sSFRs) such as those uncovered by our survey is essential for isolating the impact of young massive stellar populations from other processes.

This initial study confirms that the present body of known XMPs is incomplete \citep[as previously shown by][]{SanchezAlmeida2017}, especially at very high specific star formation rates.
SDSS photometry allowed us to select systems down to $\uband{}\sim21$, revealing that a substantial number of XMPs dominated by young stellar populations reside just below the completeness limits of large spectroscopic surveys.
Deeper HSC-SSP photometry allowed us to identify J0845+0131, an isolated and compact system with stellar mass $\lesssim 10^5 M_\odot$ at particularly low metallicity $12+\log\mathrm{O/H}=7.30\pm0.13$ and with very strong nebular line emission (H$\beta$ equivalent width $296\pm 63$ \AA{}) suggestive of individual super star clusters in SBS 0335-052E \citep[e.g.][]{Reines2008} but too faint for SDSS to resolve.
Many more such systems likely await discovery at these faint magnitudes.

Deep photometric surveys in-progress or coming online in the next decade will likely significantly expand the number and diversity of known galaxies at the lowest metallicities.
We have demonstrated that the Hyper Suprime-Cam Subaru Strategic Program \citep[HSC-SSP;][]{Aihara2018} can already be applied to reach fainter magnitudes than SDSS.
The Dark Energy Survey \citep[DES;][]{Abbott2018} and the imminent Large Synoptic Sky Telescope \citep[LSST;][]{Ivezic2008} provide access to new areas of sky in the Southern Hemisphere, and will eventually reach extremely faint objects ($\rband{}\sim 27.5$).
Color and morphological selection criteria applied to these datasets along with deep spectroscopic follow-up will likely yield a substantial number of XMPs with high equivalent-width nebular emission powered by a variety of stellar populations, bringing us closer to a complete empirical picture of extremely metal-poor stellar populations.

\section*{Acknowledgements}

We are grateful to Allison Strom for insightful conversations, and Matthew Auger for sharing his Red Channel reduction pipeline.
We also thank the referee for a thorough report which helped clarify and improve the manuscript. 
Observations reported here were obtained at the MMT Observatory, a joint facility of the University of Arizona and the Smithsonian Institution.
We thank the maintenance and administrative staff of both the MMT and Steward Observatory without whose labor this research could not be conducted, especially those of sociocultural groups traditionally excluded from academic spaces.

DPS acknowledges support from the National Science Foundation through the grant AST-1410155.
This research made use of Astropy, a community-developed core Python package for Astronomy \citep{AstropyCollaboration2013}; Matplotlib \citep{Hunter2007}; NumPy and SciPy \citep{Jones2001}; the SIMBAD database, operated at CDS, Strasbourg, France; and NASA's Astrophysics Data System.

\bibliographystyle{mnras}
\bibliography{uvgals}

\label{lastpage}
\end{document}